\documentclass[review,12pt]{elsarticle}

\usepackage{amssymb}
\usepackage{multirow}
\usepackage{booktabs}
\usepackage{makecell}
\usepackage{natbib}
\usepackage{ulem} 
\usepackage[section]{placeins} 
\usepackage{lineno,hyperref}
\usepackage{algorithm} 
\usepackage{algorithmic}



\usepackage{listings}
\usepackage{xcolor}
\definecolor{codegreen}{rgb}{0,0.6,0}
\definecolor{codegray}{rgb}{0.5,0.5,0.5}
\definecolor{codepurple}{rgb}{0.58,0,0.82}
\definecolor{backcolour}{rgb}{0.95,0.95,0.92}
\lstdefinestyle{mystyle}{
    backgroundcolor=\color{backcolour},   
    commentstyle=\color{codegreen},
    keywordstyle=\color{magenta},
    numberstyle=\tiny\color{codegray},
    stringstyle=\color{codepurple},
    basicstyle=\ttfamily\tiny,
    breakatwhitespace=false,         
    breaklines=true,                 
    captionpos=t,                    
    keepspaces=true,                 
    numbers=left,
    frame=lines,
    numbersep=3pt,                  
    showspaces=false,                
    showstringspaces=false,
    showtabs=false,                  
    tabsize=4
}
\journal{TBD}

\begin{document}

\begin{frontmatter}



\title{Linear-time online visibility graph transformation algorithm: for both natural and horizontal visibility criteria}


\author[l1]{Yusheng Huang}
\author[l1]{Yong Deng\corref{cor1}}
\ead{dengentropy@uestc.edu.cn}

\cortext[cor1]{Corresponding author.}

\affiliation[l1]{organization={Institute of Fundamental and Frontier Science, University of Electronic Science and Technology of China},
city={Chengdu},
state={Sichuan},
postcode={610054}, 
country={China}}

\begin{abstract}
Visibility graph (VG) transformation is a technique used to convert a time series into a graph based on specific visibility criteria. It has attracted increasing interest in the fields of time series analysis, forecasting, and classification. Optimizing the VG transformation algorithm to accelerate the process is a critical aspect of VG-related research, as it enhances the applicability of VG transformation in latency-sensitive areas and conserves computational resources. In the real world, many time series are presented in the form of data streams. Despite the proposal of the concept of VG's online functionality, previous studies have not thoroughly explored the acceleration of VG transformation by leveraging the characteristics of data streams. In this paper, we propose that an efficient online VG algorithm should adhere to two criteria and develop a linear-time method, termed the LOT framework, for both natural and horizontal visibility graph transformations in data stream scenarios. Experiments are conducted on two datasets, comparing our approach with five existing methods as baselines. The results demonstrate the validity and promising computational efficiency of our framework.
\end{abstract}


\begin{highlights}
\item First $O(N)$ algorithm for NVG transformation in data stream scenario. 
\item An $O(N)$ algorithm for HVG transformation in data stream scenario.
\item Our algorithms demonstrate a stat-of-the-art performance in experiments.
\end{highlights}

\begin{keyword}
Horizontal visibility graph \sep Natural visibility graph \sep Time series \sep Data stream \sep Algorithm


\end{keyword}

\end{frontmatter}


\section{Introduction}
\label{sec.intro}
The Visibility Graph (VG) transformation, as proposed by \citet{GG_NVG}, aims to construct a graph from a given time series. During the construction of a VG, certain visibility criteria must be adhered to in order to determine the adjacency relations of the graph nodes. Various visibility criteria yield different VGs. For instance, the Natural Visibility Graph (NVG) \cite{GG_NVG} is constructed using the natural visibility criterion, whereas the Horizontal Visibility Graph (HVG) \cite{GG_HVG} employs the horizontal visibility criterion. In addition to these major types of VGs, supplementary variations such as the Parametric Natural Visibility Graph \cite{GG_PNVG} and the Limited Penetrable Horizontal Visibility Graph family \cite{GG_LPHVG, GG_LPHVGF} have been introduced to enrich the VG community. A recent survey on the VG family offers further information on this topic \cite{survey_VG}. In this paper, our primary focus lies on the NVG and HVG.

The topological characteristics of VGs vary depending on the type of time series provided. For example, the NVG of a fractal time series generates a scale-free network with a power-law degree distribution \cite{GG_NVG}, whereas the HVG of a random time series produces a small-world complex network featuring a universal exponential degree distribution \cite{GG_HVG}. Other VGs possess more intricate topological properties \cite{GG_LPHVGF}. These intriguing attributes offer researchers an alternative perspective for time series related studies.

Consequently, a rapid growth in VG-related time series research has been observed over the past decade. Some studies emphasize the analysis of time series through the application of VG transformation. For instance, \citet{exa_analysis1} introduced a VG-based system to analyze the global fuel market, offering insights from complex network perspectives; \citet{exa_analysis2} employed VG technology in log file analysis, highlighting its future potential in reverse engineering and software anomaly detection; \citet{exa_analysis3} devised a VG-based multi-scale network method for the analysis of medical series, among others.

Alternatively, other research focuses on practical applications, such as time series forecasting, classification, and anomaly detection. For example, \citet{exa_forecast1} developed a VG-based multi-sub-graph method for time series forecasting; \cite{exa_forecast2} integrated VG technology with a random walk approach to predict time series; \citet{exa_forecast3} proposed an innovative neural network structure inspired by VG for time series classification.

In real-world scenarios, many time series take the form of data streams, e.g. financial time series like stock price, nature-related time series such as temperature or wind speed series, and medical time series like blood pressure, among others. For VG-related research of time series,  a crucial step is to transform the time series to VG using VG transformation algorithms. Therefore, the ability to generate VG from the data stream at regular intervals\cite{baseline_MS}, or the so called online functionality\cite{baseline_BST} of the VG algorithm, is of great importance. Accelerating the speed of the VG transformation can enable its application in low-latency demanding areas like high-frequency trading systems, online recommendation systems\cite{exa_recommendsys}, or real-time anomaly detection software\cite{exa_anolydetect}, etc. 

\begin{figure}[htb!]
	\centering
	\includegraphics[width=0.75\textwidth]{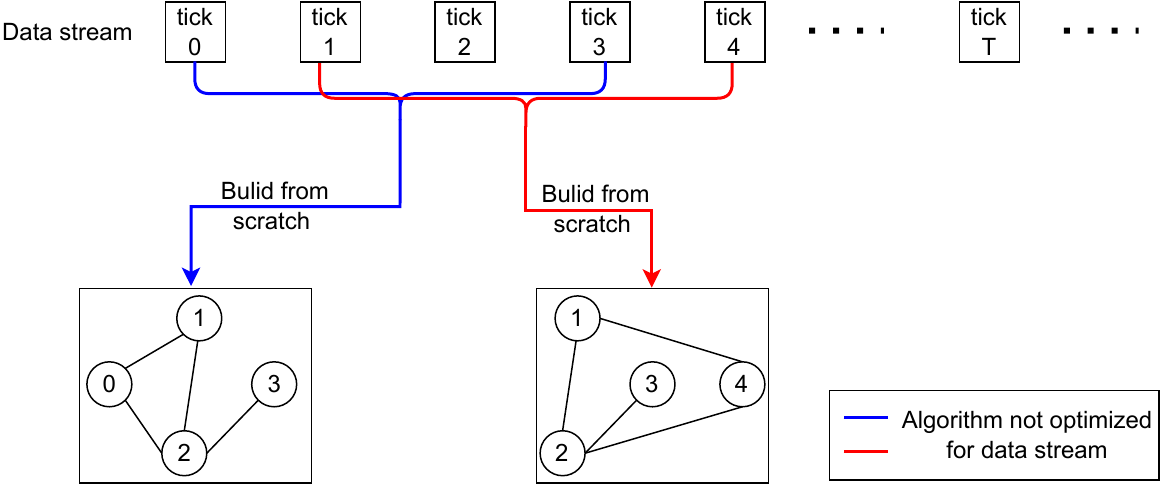}
	\caption{Build from scratch"-style VG method in data stream scenario.}
	\label{fig.datastream_ineffi}
\end{figure}

\begin{figure}[htb!]
	\centering
	\includegraphics[width=0.75\textwidth]{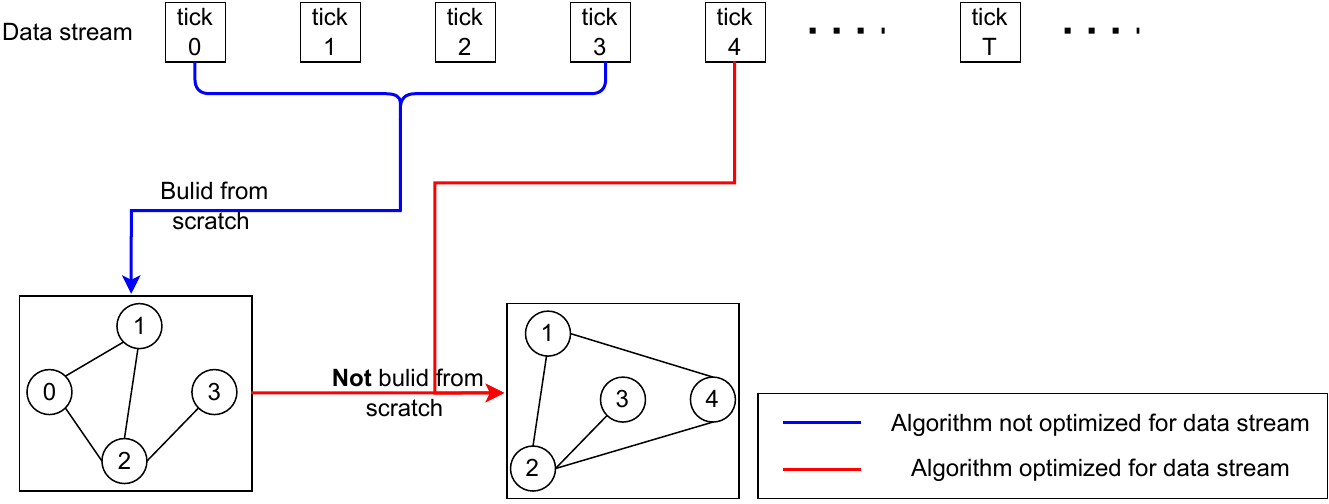}
	\caption{Efficient update and elimination"-style VG method in data stream scenario.}
	\label{fig.datastream_effi}
\end{figure}

By summing up previous articles\cite{baseline_BST, baseline_MS}, we propose that an efficient online VG algorithm should have the features\footnote{Would be quoted by "the standards of VG's online functionality" in the rest of our paper.} listed below:
\begin{itemize}
    \item Efficient update\cite{baseline_BST, baseline_MS}: With $(s_0, s_1, \cdots, s_T)$ as the whole series, $T$ refers to the latest time point and $N$ denotes the interval size, $(s_{T - N + 1}, \cdots, s_{T})$ is the time series of current interest. The online VG algorithm should build the $VG_{nodes=(T - N + 1, \cdots, T)}$ of current interval $(s_{T - N + 1}, \cdots, s_{T})$ by fully utilizing the generated $VG_{nodes=(T - N, \cdots, T - 1)}$ at the last interval $(s_{T - N}, \cdots, s_{T - 1})$ instead of building it from scratch. During the efficient update, in-place update instead of update-after-copy over the adjacent data structure is preferred to reduce both computation and memory resources.
    \item Efficient elimination: As in-place update is conducted during efficient update, the resulted VG would become $VG_{nodes=(T - N, T - N + 1, \cdots, T)}$ with $node=(T - N)$ lying out the current interval $(s_{T - N + 1}, \cdots, s_{T})$. Therefore, in-place elimination of the undesired $node=(T - N)$ in $VG_{nodes=(T - N, T - N + 1, \cdots, T)}$ is required to build the final $VG_{nodes=(T - N + 1, \cdots, T)}$.
\end{itemize}
Fig. \ref{fig.datastream_ineffi} and Fig. \ref{fig.datastream_effi} show the difference between the "efficient update and elimination"-style method and the "build from scratch"-type method. 

However, previous research either fails to fulfill the above standards, suffers from high time and space complexity, or lacks generalization for both NVG and HVG (Detailed analysis of related works is provided in Sec. \ref{sec.related}). This motivates us to propose the LOT (linear-time online transformation for visibility graph) framework. The LOT framework uses the adjacent dict with node set to store the adjacent relation of the graph, which allows efficient elimination of the node that lies out of the current interval but lies in the previous interval. The LOT framework uses the "traverse from back" strategy for visibility check and edge construction, which results in the worse-case time complexity of $O(N)$ and the auxiliary space complexity of $O(1)$. Our LOT framework works for both NVG and HVG: the only difference between the proposed LOT-NVG algorithm and the proposed LOT-HVG algorithm is the visibility checking function. For HVG transformation, we use the monotonic stack idea of article\cite{baseline_MS} to optimize the "traverse from back" strategy of the LOT framework and proposed the LOT-HVG\_MSopt algorithm.

The contributions of this paper are:
\begin{enumerate}
    \item The proposed concept of the features of VG's online functionality, of which the in-place update and elimination are under-emphasized in the previous research.
    \item The proposed LOT framework and the corresponding implementations for NVG (LOT-NVG) and HVG (LOT-HVG and LOT-HVG\_MSopt). The worst-case time complexity of the three implementations is $O(N)$. As for the auxiliary space complexity, it is $O(1)$ for LOT-NVG and LOT-HVG algorithms and is $O(N)$ for the LOT-HVG\_MSopt method.
    \item To the best of our knowledge, experiments show that the proposed algorithms achieve the current stat-of-the-art performance. 
\end{enumerate}

The rest of the paper is structured as follows: Sec. \ref{sec.pre} provides some preliminaries about the concept of VG and discusses the related works of VG transformation algorithm; Sec. \ref{sec.proposed} illustrates the proposed LOT framework; Sec. \ref{sec.exp} demonstrates the experimental results and Sec. \ref{sec.conclu} is the conclusion.

\section{Preliminaries and Related works}
\label{sec.pre}
In this section, we first provide some preliminaries about the concept of NVG and HVG, and the corresponding transformation algorithms. The related works of NVG and HVG's transformation algorithms are presented in Sec. \ref{sec.related}.
\subsection{Preliminaries}

As is described in Sec. \ref{sec.intro}, the visibility graph transformation is to transform a given time series to a VG following certain visibility criteria.

\begin{figure}[htb!]
	\centering
	\includegraphics[width=0.75\textwidth]{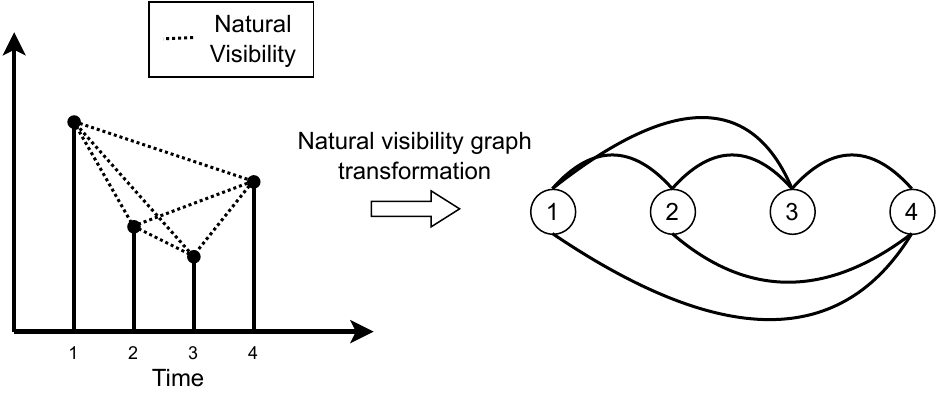}
	\caption{Example of the NVG transformation.}
	\label{fig.NVG}
\end{figure}

\begin{figure}[htb!]
	\centering
	\includegraphics[width=0.75\textwidth]{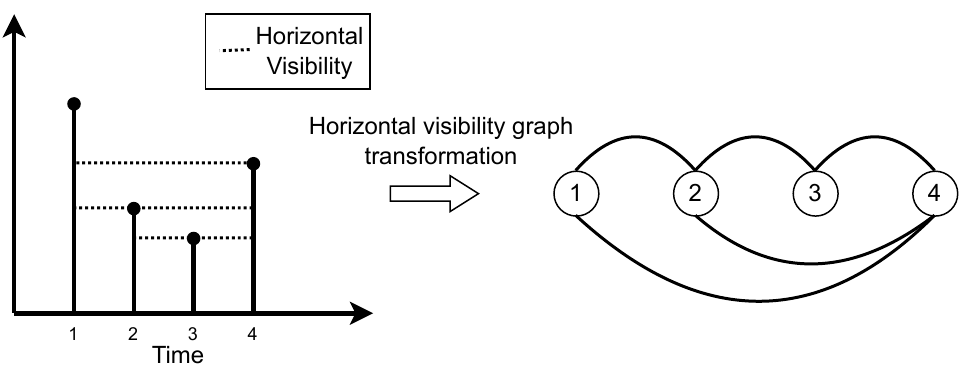}
	\caption{Example of the HVG transformation.}
	\label{fig.HVG}
\end{figure}

\noindent\textbf{Natural visibility criterion\cite{GG_NVG}:} given a time series $(s_{T - N + 1}, \cdots, s_{T})$, and the corresponding NVG $NVG_{nodes=(T - N + 1, \cdots, T)}$, for any tow  nodes $node=(i)$ and $node=(j)$ ($T - N + 1 \leq i \leq j \leq T$) in $NVG_{nodes=(T - N + 1, \cdots, T)}$, the edge $edge=(ij)$ between the two nodes exits if and only if
\begin{equation}
    s_{k} < s_i + (s_j - s_i)\frac{k - i}{j - i},\ \forall\ i \leq k \leq j
\end{equation} holds.

\noindent\textbf{Horizontal visibility criterion\cite{GG_HVG}:} given a time series $(s_{T - N + 1}, \cdots, s_{T})$, and the corresponding HVG $HVG_{nodes=(T - N + 1, \cdots, T)}$, for any tow  nodes $node=(i)$ and $node=(j)$ ($T - N + 1 \leq i \leq j \leq T$) in $HVG_{nodes=(T - N + 1, \cdots, T)}$, the edge $edge=(ij)$ between the two nodes exits if and only if
\begin{equation}
    s_{k} < min(s_{i}, s_{j}),\ \forall\ i \leq k \leq j
\end{equation} holds.

Fig. \ref{fig.NVG} and \ref{fig.HVG} illustrate the NVG and HVG transformations, respectively. In the left portion of Fig. \ref{fig.NVG}, the dashed lines represent the natural visibility between the time ticks; consequently, the corresponding nodes in the NVG (the right section of Fig. \ref{fig.NVG}) are connected through these natural visibility relations. Similarly, the HVG transformation is depicted in \ref{fig.HVG}.

\begin{lstlisting}[language=Python, firstline=1, lastline=9, label=lst.basic, caption=The basic VG algorithm.]
# Python-style code snippet for demonstration.
def basic_algo(series, ticks):
    adj_list = []
    left_index = 0
    right_index = len(ticks) - 1
    for target_index in range(len(ticks)):
        (node_ID, node_visible) = visibility_checking_function(series, ticks, target_index, left_index, right_index)
        adj_list.append([node_ID, node_visible])
    return adj_list
\end{lstlisting}

\noindent\textbf{The Basic algorithm\cite{GG_NVG,GG_HVG, baseline_basic}} Given series $(s_{T - N + 1}, \cdots, s_{T})$, the original transformation algorithm consists of two loops. In the outer loop, each tick $s_{i}, \ T - N + 1\leq i\leq T$ are traversed (line 6 to 8 in Listing \ref{lst.basic}).  Within the inner loop, the visibility relations between tick $s_{i}$ and the series $(s_{T - N + 1}, \cdots, s_{i - 1})$ to its left are examined (line 8 to 16 in Listing \ref{lst.NVG} or line 8 to 15 in Listing \ref{lst.HVG}); the same procedure is applied to tick $s_{i}$ and the series $(s_{i + 1}, \cdots, s_{T})$ to its right (line 18 to 26 in Listing \ref{lst.NVG} or line 17 to 24 in Listing \ref{lst.HVG}). The basic algorithm\cite{baseline_basic} (generalized for both natural visibility and horizontal visibility transformation) possesses a time complexity of $O(N^2)$, independent of the input time series. For a detailed illustration, refer to Listings \ref{lst.basic}, \ref{lst.NVG}, and \ref{lst.HVG}.

\subsection{Related works}
\label{sec.related}

\begin{table}[htb]
\centering
\caption{Brief summary of the visibility transformation algorithms in data stream scenario.}
\resizebox{0.95\textwidth}{23mm}{
\begin{tabular}{cccccc}
\toprule[0.75pt]
\makecell{Auxiliary \\ Space Complexity} & \makecell{Auxiliary \\ Data Structure} & Method & \makecell{Worse Case \\ Time Complexity} & \makecell{Data Structure for \\ Adjacent Relationship} & \makecell{For both \\ NVG and HVG?}\\
\midrule[0.3pt]
 $*^a$& $*^a$ & SC\cite{baseline_SC} & $O(N^2)$ & Dict$^b$ & Yes\\
 \midrule[0.3pt]
\multirow{3}{*}{$O(1)$} & - & Basic\cite{baseline_basic} & $O(N^2)$ & List$^b$ & Yes\\
 & - & DC\cite{baseline_DC} & $O(N^2)$ & List$^b$ & Yes\\
 & - & \textbf{LOT(Ours)} & $O(N)$ & Dict & Yes\\
\midrule[0.3pt]
\multirow{3}{*}{$O(N)$} & Binary search tree & BST\cite{baseline_BST} & $O(N^2)$ & List$^b$ & Yes\\
 & Monotonic stack & LT\cite{baseline_MS} & $O(N + E)$ & List$^b$ & Only for HVG\\
 & Monotonic stack & \textbf{LOT-HVG\_MSopt(Ours)} & $O(N)$ & Dict & Only for HVG\\
\bottomrule[0.75pt]
\end{tabular}
\label{table.summary}
}
\footnotesize{a.Depends on the adopted sort method.\\\vspace{-3mm}b.Inferred from the corresponding references or source codes.}
\end{table}

\textbf{DC method\cite{baseline_DC}:} In 2015, \citet{baseline_DC} made the initial attempt to reduce the time complexity of the basic algorithm\cite{baseline_basic} by using the "divide and conquer" strategy. Given series $(s_{T - N + 1}, \cdots, s_{T})$, at each step of the recursion: 1) the DC method\cite{baseline_DC} first finds the highest value $s_{hightest}\ (T - N + 1 <= highest <= T)$ of the series; 2) the DC method\cite{baseline_DC} checks the visibility relation between $Node_{hightest}$ and other nodes; 3) the given series is split into two sub-series $(s_{T - N + 1}, \cdots, s_{hightest - 1})$ and $(s_{hightest + 1}, \cdots, s_{T})$, and then the sub-series are passed to the next step of recursion respectively. The time complexity of the DC method\cite{baseline_DC} is $O(NlogN)$ for randomly generated or Conway series; while for monotonous or bowel-like series, it is $O(N^2)$\cite{baseline_basic}. The DC method\cite{baseline_DC} works for both NVG and HVG.

\textbf{SC method\cite{baseline_SC}:} Article\cite{baseline_SC} pointed out that the recursion strategy of the DC method\cite{baseline_DC} would result in low memory efficiency in some programming languages, and proposed the SC ("sort and conquer") algorithm\cite{baseline_SC} to address this issue. Given series $(s_{T - N + 1}, \cdots, s_{T})$: 1) the SC method first sort the series by descending order; 2) in the outer loop, each node $Node_{i}$ in the sorted series is traversed; 3) in the inner loop, the neighbors of node $Node_{i}$ are traversed to determined which nodes the visibility check of node $Node_{i}$ should be conducted on. The time complexity of the SC method\cite{baseline_SC} is $O(NlogN)$ for well-balanced series like white noise; while for extremely unbalanced series, it is $O(N^2)$\cite{baseline_SC}. The SC method\cite{baseline_SC} works for both NVG and HVG.

\textbf{BST method\cite{baseline_BST}:} In \cite{baseline_BST}, the authors emphasize the importance of the online functionality of the VG algorithm and propose the BST method\cite{baseline_BST}. The BST method\cite{baseline_BST} utilizes the binary search tree to encode the visibility relationship between the nodes and supports the efficient merging of two VGs. The time complexity of the BST method\cite{baseline_BST} is $O(NlogN)$ for input series which results in a balanced binary search tree; while for the worse case, it is $O(N^2)$\cite{baseline_BST}. The BST method\cite{baseline_BST} works for both NVG and HVG.

\textbf{LT method\cite{baseline_MS}:} To further enhance the online functionality of the HVG algorithm, the authors proposed the LT method\cite{baseline_MS} with a simpler data structure. The LT method\cite{baseline_MS} adopts the monotonic stack to narrow the scope of the visibility check. The time complexity of the LT method\cite{baseline_MS} is $O(N)$ regardless of the types of inputs. The LT method\cite{baseline_MS} only works for the HVG. In \cite{baseline_MS}, the LT method is claimed to be the current state-of-the-art in accelerating the HVG transformation.

\textbf{Other methods:} The DTHVG method\cite{baseline_DTHVG} is also developed to accelerate the HVG transformation; but its performance is inferior to the LT method\cite{baseline_MS} according to\cite{baseline_MS}; therefore we omit the detailed discussion. In \cite{baseline_others_sliding}, the concept of "sliding window visibility graph" is proposed to accelerate the visibility transformation; however, the output graph is neither the original NVG nor the original HVG; hence this method fills out of the scope of our paper.

From the perspective of the VG's online functionality, we sum up the aforementioned algorithms:
\begin{itemize}
    \item The DC method\cite{baseline_DC} and the SC method\cite{baseline_SC} do not fulfill the "efficient update" and "efficient elimination" standards (proposed by us in Sec. \ref{sec.intro}), as they are not designed for streaming data and need to build the VG from scratch when new data points arrived. The worst-case time complexity of them in a data stream scenario is $O(N^2)$.
    \item The binary search tree data structure of the BST method\cite{baseline_BST} supports efficient update, but the authors do not provide a solution for efficient elimination. Hence, the BST method\cite{baseline_BST} also needs to build the VG from scratch when new data points arrive, which costs a $O(N^2)$ worst-case time complexity.
    \item The LT method\cite{baseline_MS} is capable of efficient update. However, the adopted adjacent list results in a $O(E)$ time complexity during the node elimination as all edges in the graph need to be traversed. Hence, the overall time complexity of the LT method\cite{baseline_MS} in the data stream scenario is $O(N + E)$.
\end{itemize}
A brief summary of the above methods is presented in Table \ref{table.summary}.

\section{The proposed algorithm}
\label{sec.proposed}

\begin{figure}[htb!]
	\centering
	\includegraphics[width=1\textwidth]{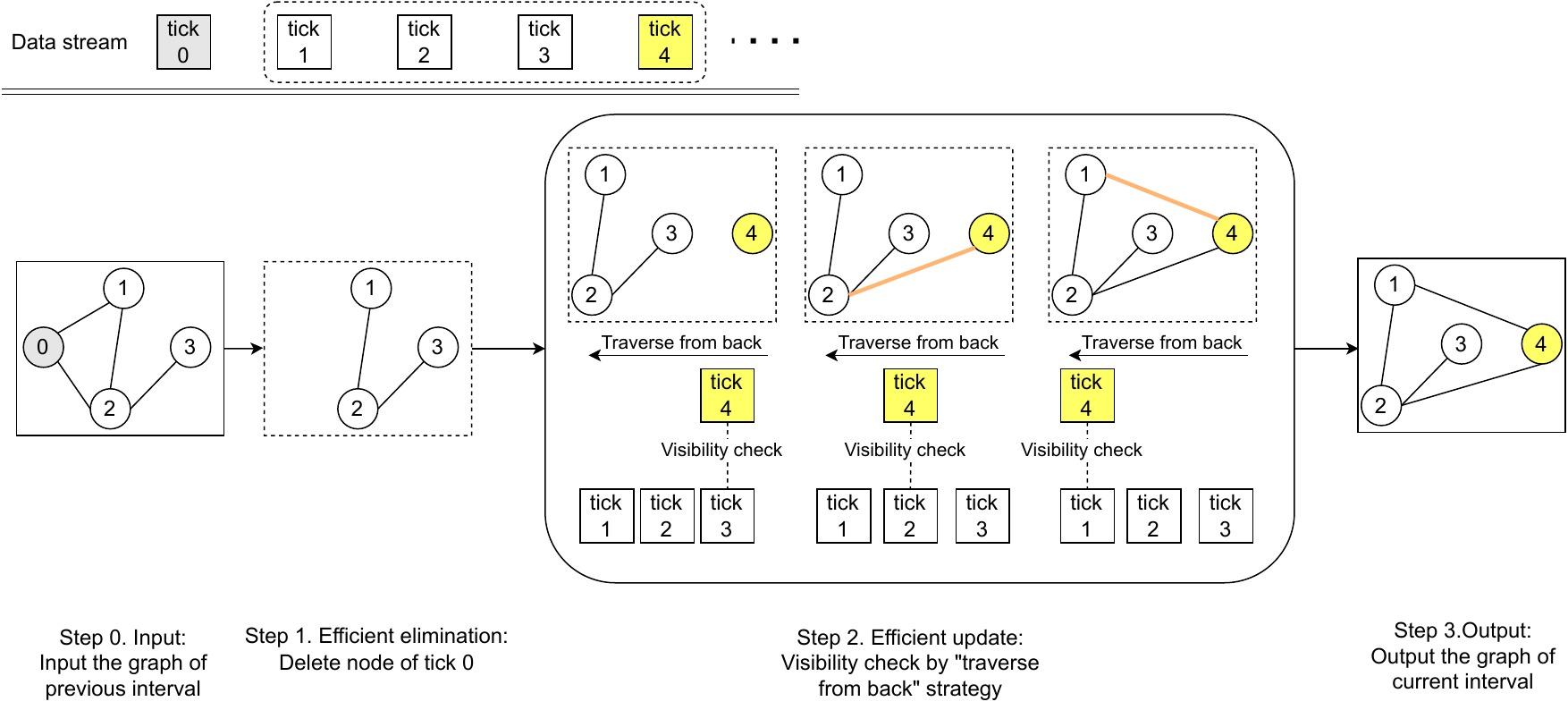}
	\caption{The proposed LOT framework.}
	\label{fig.LOT}
\end{figure}

In this section, the proposed LOT framework is demonstrated\footnote{For the convenience of illustration, the code snippets of the LOT framework are provided in Listing \ref{lst.LOT} in Sec. \ref{sec.proposed}, and Listing \ref{lst.LOT-NVG} and \ref{lst.LOT-HVG} in \ref{app.code}.}. A visual representation of the LOT framework is provided in Fig. \ref{fig.LOT}.

\begin{lstlisting}[language=Python, firstline=1, lastline=26, label=lst.LOT, caption=LOT framework.]
# Python-style code snippet for demonstration.
from collections import defaultdict
from typing import Union

def LOT(adj_dict: Union[defaultdict, None], ori_series: list, index_left: int, index_right: int) -> defaultdict:
    if index_left == 0 or adj_dict == None:
        # In the very first start, build the graph from scratch using any VG algorithm.
        adj_dict = defaultdict(set)
        adj_dict = build_from_scratch(adj_dict, ori_series, index_left, index_right)
        return adj_dict
    else:
        # Online mode.

        # Efficient elimination.
        if index_left - 1 in adj_dict:
            del adj_dict[index_left - 1]
        node_list = list(adj_dict.keys())
        for node_index in node_list:
            if index_left - 1 in adj_dict[node_index]:
                adj_dict[node_index].remove(index_left - 1)
            if len(adj_dict[node_index]) == 0:
                del adj_dict[node_index]

        # Efficient update.
        adj_dict = efficient_update(adj_dict, ori_series, index_left, index_right)
        return adj_dict
\end{lstlisting}

In the majority of previous studies, the data structure of the VG is an adjacent list with edge pairs as elements. This implies that to execute node elimination, all the edge pairs in the adjacent list must be traversed, resulting in a computational complexity of $O(E)$ ($E$ is the number of edges in the VG). In the LOT framework, we employ an adjacency dict, with the node index as the key and a set containing the node's neighbors as the corresponding value. This special adjacent dict serves as the foundation for efficient elimination, as the edges between $node=(T - N)$ and $nodes=(T - N + 1, \cdots, T - 1)$ could be deleted in $O(N)$.

Upon the arrival of the first data stream interval $(s_{0}, \cdots, s_{N - 1})$, the VG must be constructed from scratch, and in this boundary case, any existing VG algorithms are suitable (line 6 to 10 in Listing \ref{lst.LOT}). Once the VG of the first interval has been built, the LOT algorithm works in the online mode, processing the most recent interval $(s_{T - N + 1}, \cdots, s_{T})$ in real-time:
\begin{enumerate}[Step 1.]
    \item Efficient elimination (line 14 to 22 in Listing \ref{lst.LOT}): given the VG from the previous interval $VG_{nodes=(T - N, \cdots, T - 1)}$, we eliminate the node that does not belong to the current interval. With the assistance of the specially designed adjacent dict, this process can be completed in $O(N)$.
    \item Efficient update (line 24 to 25 in Listing \ref{lst.LOT}): The "traverse from back" strategy is employed to examine the visibility relations between tick $s_{T}$ and the series $(s_{T - N + 1}, \cdots, s_{T - 1})$ to its left. The detailed procedures are provided in Listing \ref{lst.LOT-NVG} for the NVG and Listing \ref{lst.LOT-HVG} for the HVG. As all nodes are only traversed once in this process, the time complexity is $O(N)$.
    \item Output the generated VG.
\end{enumerate}

\noindent\textbf{Time and space complexity:} As is demonstrated in the previous paragraph, the time complexity of both efficient elimination and efficient update is both $O(N)$. Consequently, the overall time complexity of the LOT framework is $O(N)$. The LOT framework does not use any auxiliary data structure, leading to an $O(1)$ auxiliary space complexity.

\noindent\textbf{The Correctness of the LOT framework:} \begin{itemize}
    \item During the efficient elimination, given the previous VG $VG_{nodes=(T - N, \cdots, T - 1)}$, $node=(T - N)$ and its connecting edges are removed, which does not affect either the natural visibility ($s_{k} < s_i + (s_j - s_i)\frac{k - i}{j - i},\ \forall (i,k,j)\ where\ T - N + 1\leq i \leq k \leq j \leq T - 1$) or the horizontal visibility ($s_{k} < min(s_{i}, s_{j}),\ \forall (i,k,j)\ where\ T - N + 1\leq i \leq k \leq j \leq T - 1$) relations among the other nodes $nodes=(T - N, \cdots, T - 1)$. This process produces a temporary output $NVG_{nodes=(T - N + 1, \cdots, T - 1)}$ for the NVG transformation or $HVG_{nodes=(T - N + 1, \cdots, T - 1)}$ for the HVG transformation.
    \item Given $NVG_{nodes=(T - N + 1, \cdots, T - 1)}$, for the NVG, the minimum slope $slope_{minimum}$ (initialized to positive infinite) is kept during the procedures of the "traverse from back" strategy. During the traverse loop, the current slope $slope_{i,T}$ between the current tick $i$ and the latest tick $T$ is calculated. If $slope_{i,T} \leq slope_{minimum} = min(slop_{i+1,T}, slop_{i+2,T}, \cdots slop_{T-1,T})$, the natural visibility exists between tick $i$ and $T$ according to the natural visibility criteria. Hence, the algorithm could generate the correct NVG by adding an edge between tick $i$ and $T$. The final output is $NVG_{nodes=(T - N + 1, \cdots, T)}$. Refer to Listing \ref{lst.LOT-NVG} for more details.
    \item Given $HVG_{nodes=(T - N + 1, \cdots, T - 1)}$, for the HVG, the maximum value $s_{maximum}$ (initialized to negative infinite) is maintained during the procedures of the "traverse from back" strategy. During the traverse loop, the current value $s_{i}$, the value $s_T$ of the latest tick $T$, and the maximum value $s_{maximum}$ are compared. If $min(s_{i}, s_T) > s_{maximum} = max(s_{i+1}, s_{i+2}, \cdots s_{T-1})$, the horizontal visibility exists between tick $i$ and $T$ according to the horizontal visibility criteria. Hence, the algorithm could generate the correct HVG by adding an edge between tick $i$ and $T$. The final output is $HVG_{nodes=(T - N + 1, \cdots, T)}$. Refer to Listing \ref{lst.LOT-HVG} for more details.
\end{itemize}

\section{Experiment}
\label{sec.exp}
\subsection{Settings}
\begin{figure}[htb!]
	\centering
	\includegraphics[width=1\textwidth]{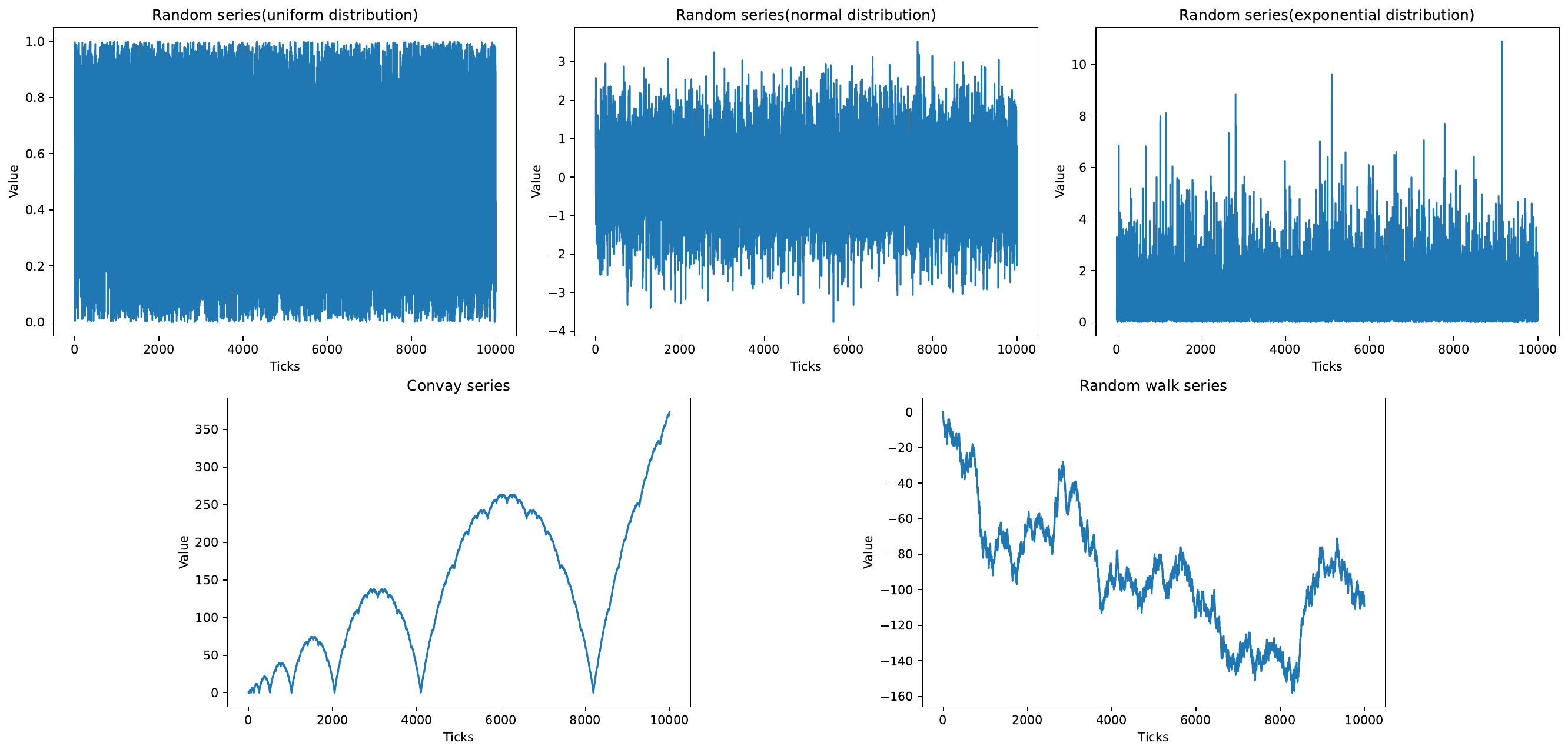}
	\caption{Synthetic dataset.}
	\label{fig.synthetic_series}
\end{figure}
\begin{figure}[htb!]
	\centering
	\includegraphics[width=1\textwidth]{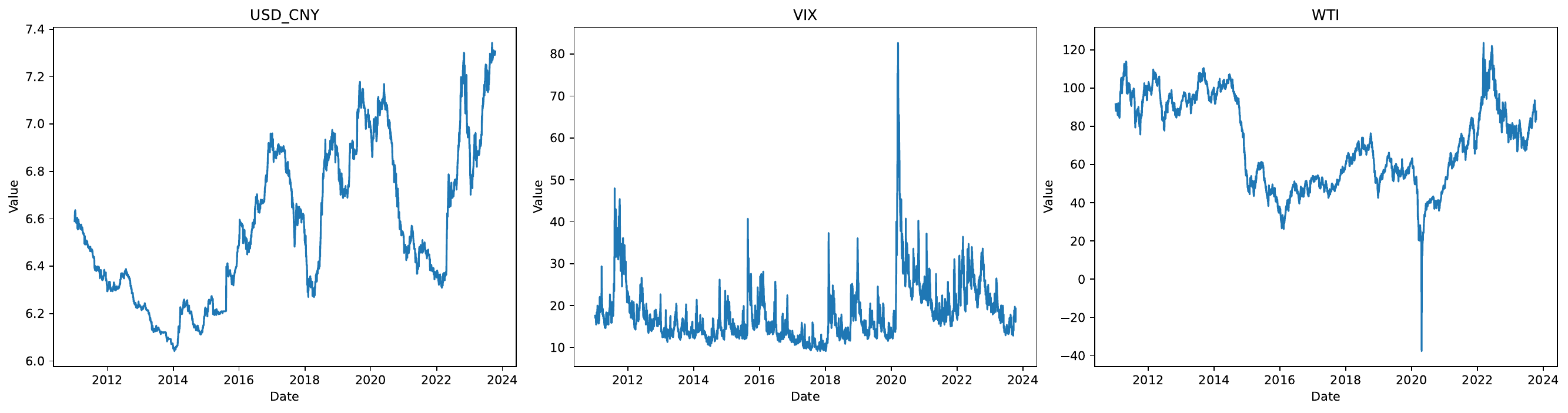}
	\caption{Real-world dataset.}
	\label{fig.realworld_series}
\end{figure}
\noindent\textbf{Synthetic dataset:} Following\cite{baseline_DC, baseline_MS}, we generate five series whose lengths are 10000 for our experiment:
\begin{itemize}
    \item a random series generated by the "random.uniform"\footnote{The random seed is set to 1024.} function of the Numpy\footnote{https://numpy.org} package with "low=0.0" and "high=1.0" as parameters;
    \item a random series generated by the "random.normal" function of the Numpy package with "loc=0.0" and "scale=1.0" as parameters;
    \item a random series generated by the "random.exponential" function of the Numpy package with "scale=1.0" as parameter;
    \item a Conway series generated following:
    \begin{equation}
        s(t) = \left\{ \begin{array}{cc}
             1 & \ ,t = 1\ or\ t = 2;  \\
             s(s(t - 1)) + s(t - s(t - 1)) & \ , t > 2.
        \end{array}\right.
    \end{equation}
    \item a random walk series generated following:
    \begin{equation}
        s(t) = \left\{ \begin{array}{cc}
             0 & \ ,t = 1;  \\
             s(t - 1) + \epsilon\ \ (\epsilon\sim\left\{p(\epsilon=-1)=0.5,\  p(\epsilon=1)=0.5\right\}) & \ , t > 1.
        \end{array}\right.
    \end{equation}
\end{itemize}
One could refer to Fig. \ref{fig.synthetic_series} for the visual illustration of the synthetic dataset.

\noindent\textbf{Real-world dataset:} We obtain three series real-world financial series from the Internet\footnote{https://cn.investing.com}:
\begin{itemize}
    \item USD\_CNY\footnote{https://cn.investing.com/currencies/usd-cny-historical-data}: The close price series of the exchange rate between USD and CNY, spanning from January 3rd, 2011 to October 13th, 2023.
    \item VIT\footnote{https://cn.investing.com/indices/volatility-s-p-500-historical-data}: The close price series of the VIT index, spanning from January 3rd, 2011 to October 13th, 2023.
    \item WTI\footnote{https://cn.investing.com/commodities/crude-oil-historical-data}: The close price series of the WTI crude oil index, spanning from January 3rd, 2011 to October 13th, 2023.
\end{itemize}

\noindent\textbf{Baselines:} For NVG transformation, the Basic-NVG\cite{baseline_basic}, the DC-NVG\cite{baseline_DC}, the SC-NVG\cite{baseline_SC}, and the BST-NVG\cite{baseline_BST} method are adopted as baseline algorithms. While for HVG transformation, the Basic-HVG\cite{baseline_basic}, the DC-HVG\cite{baseline_DC}, the SC-HVG\cite{baseline_SC}, the BST-HVG\cite{baseline_BST}, and the LT algorithm\cite{baseline_MS} are adopted as baselines.

\noindent\textbf{Experimental method:}
\begin{figure}[htb!]
	\centering
	\includegraphics[width=0.75\textwidth]{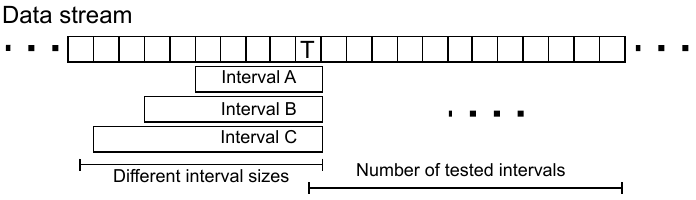}
	\caption{Experimental method.}
	\label{fig.exp}
\end{figure}
To emulate the data stream scenario, the evaluated algorithm is required to process the time series at regular intervals for 100 iterations in a moving-window fashion, as illustrated in Fig. \ref{fig.exp}. The interval sizes are set to $[10, 50, 100, 250, 500, 750, 1000, 1500, 2000]$ for synetic dataset and $[10, 50, 100]$ for real-world dataset. For the synthetic dataset, we compute the average computational time across the 100 iterations; for the real-world dataset, we report the cumulative running time of the 100 iterations. All experiments are performed five times, and the standard error is presented (refer to Fig. \ref{fig.expres_NVG_syn}, \ref{fig.expres_HVG_syn}, \ref{fig.expres_NVG_realworld}, and \ref{fig.expres_HVG_realworld}; note that some standard errors may be too narrow to be visible). When the very first interval arrives, the LOT method needs to build the VG from scratch, for which we use the DC-NVG\cite{baseline_DC} and LT\cite{baseline_MS} as the offline algorithm; after that, the LOT framework runs in online mode.

\subsection{Results for synthetic dataset}
\begin{figure}[htb!]
	\centering
	\includegraphics[width=1\textwidth]{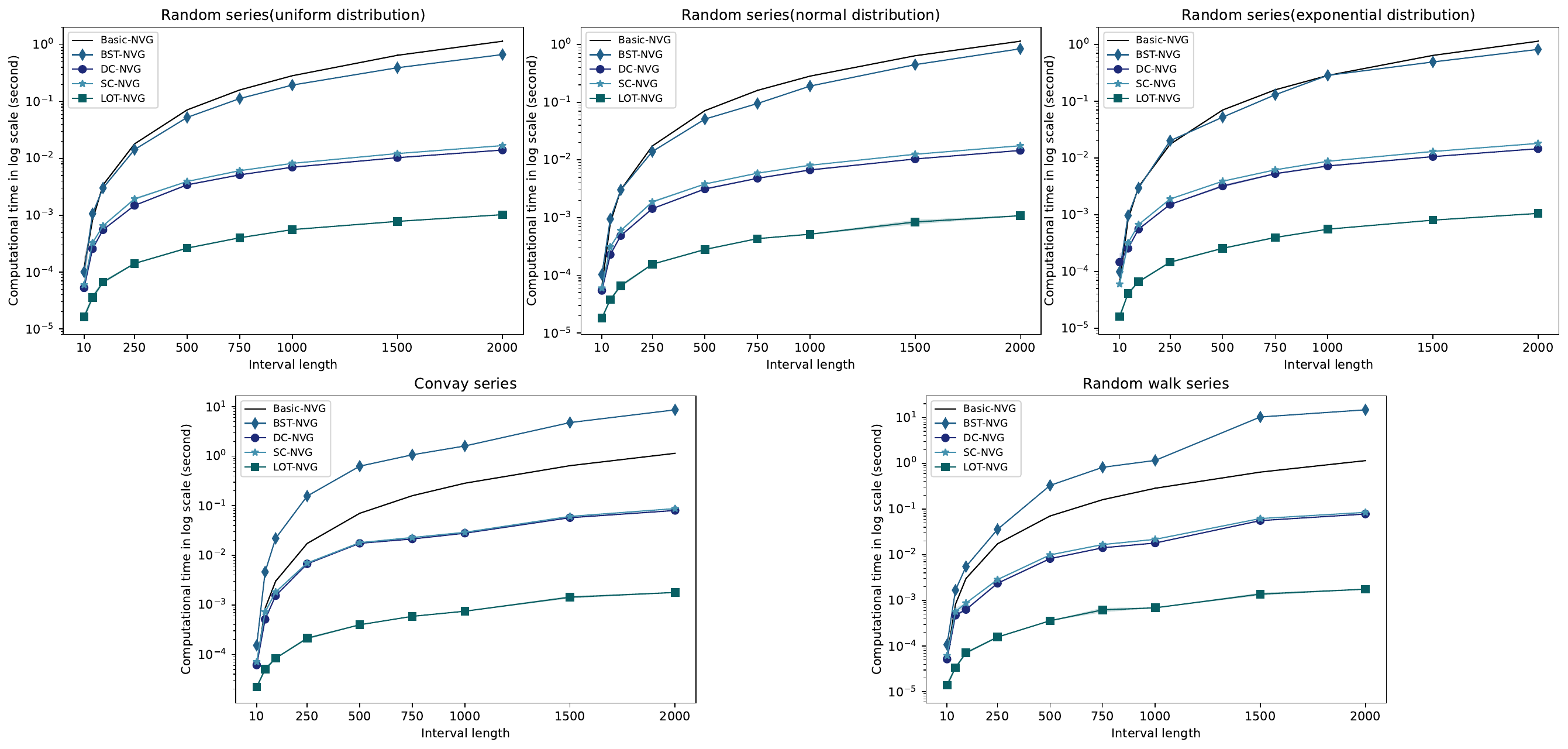}
	\caption{Average computational time for different NVG algorithms in synthetic dataset.}
	\label{fig.expres_NVG_syn}
\end{figure}
\begin{figure}[htb!]
	\centering
	\includegraphics[width=1\textwidth]{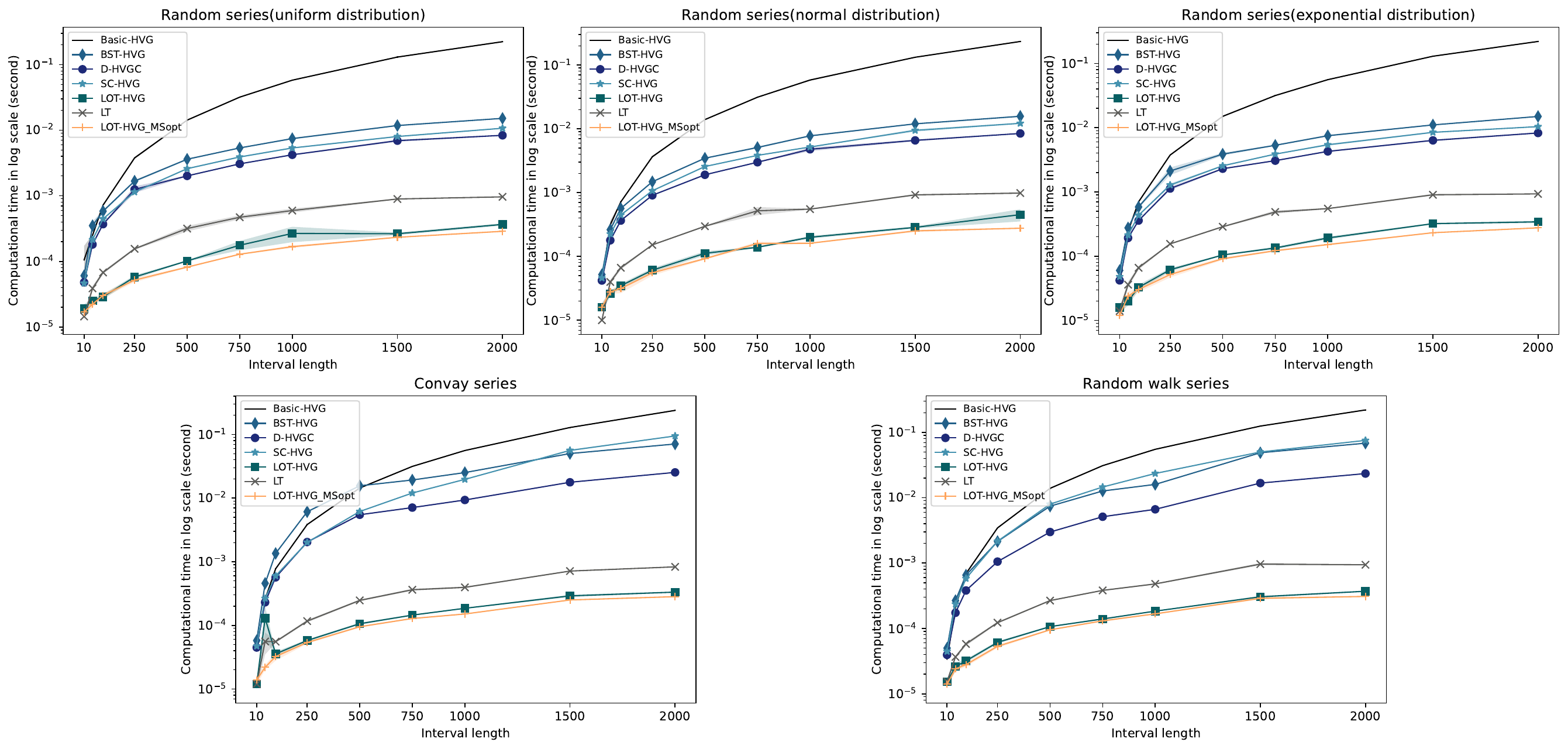}
	\caption{Average computational time for different HVG algorithms in synthetic dataset.}
	\label{fig.expres_HVG_syn}
\end{figure}
The experimental results for the NVG transformation algorithms are presented in Fig. \ref{fig.expres_NVG_syn} and Tables \ref{table.exp_NVG_syn_1}, \ref{table.exp_NVG_syn_2}, and \ref{table.exp_NVG_syn_3}. As illustrated, the proposed LOT-NVG algorithm consistently outperforms the competing methods in terms of computational efficiency across all five series, with DC-NVG\cite{baseline_DC} ranking second. The LOT-NVG algorithm takes approximately one millisecond to generate an NVG for all five series, even when the interval size reaches 2000; in contrast, the execution time of the second-fastest method, DC-NVG\cite{baseline_DC}, is nearly an order of magnitude greater than that of our algorithm.

The experimental results for the HVG transformation algorithms are provided in Fig. \ref{fig.expres_HVG_syn} and Tables \ref{table.exp_HVG_syn_1}, \ref{table.exp_HVG_syn_2}, and \ref{table.exp_HVG_syn_3}. When the interval size is 10, the LT method \cite{baseline_MS} is the fastest in two of the tested series; however, under other circumstances, the LT method \cite{baseline_MS} occupies the third-fastest position. The performance of the proposed LOT-HVG and LOT-HVG\_MSopt methods alternates in achieving the top position when the interval size is larger than 10, requiring less than half a millisecond to generate the HVG in all the tested cases. When the interval size exceeds 1000, the LOT-HVG\_MSopt performs better than the LOT-HVG across all five series, indicating that the proposed LOT framework is compatible with the monotonic stack idea\cite{baseline_MS} for HVG transformation.

\subsection{Results for real-world dataset}
\begin{figure}[htb!]
	\centering
	\includegraphics[width=1\textwidth]{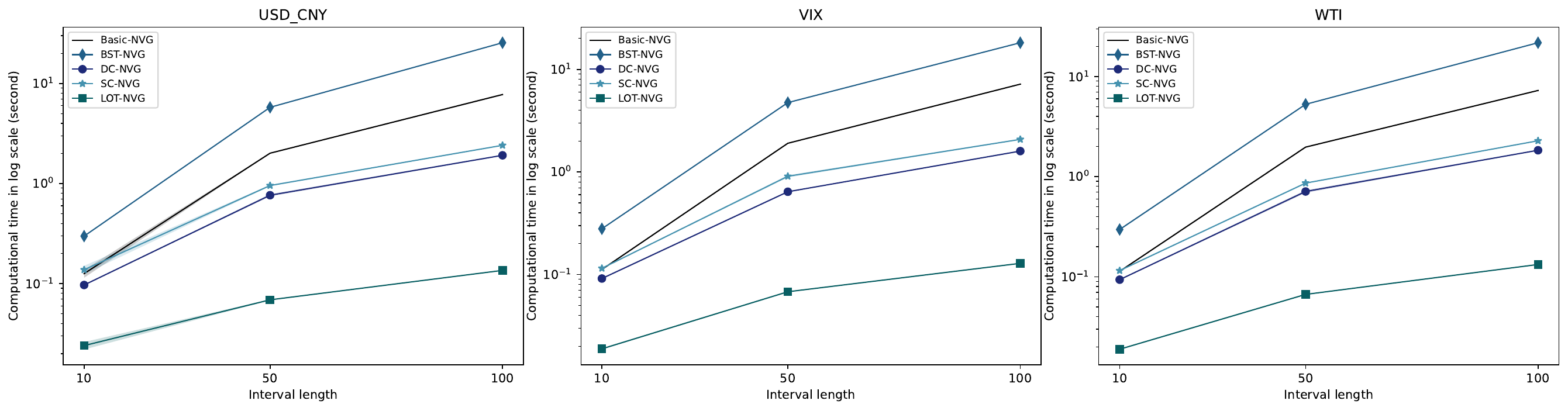}
	\caption{Total computational time for different NVG algorithms in real-world dataset.}
	\label{fig.expres_NVG_realworld}
\end{figure}
\begin{figure}[htb!]
	\centering
	\includegraphics[width=1\textwidth]{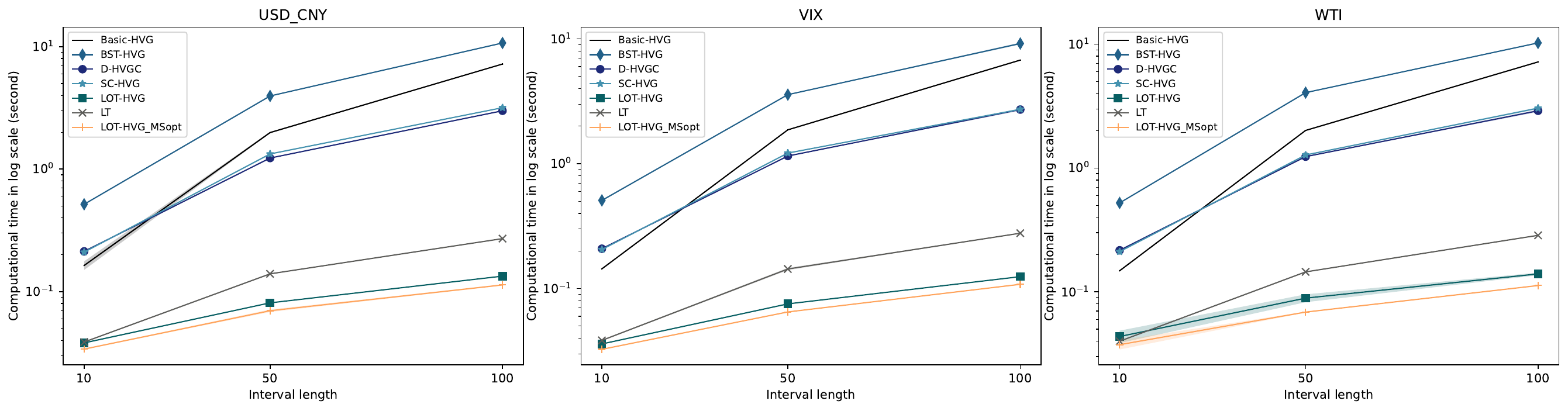}
	\caption{Total computational time for different HVG algorithms in real-world dataset.}
	\label{fig.expres_HVG_realworld}
\end{figure} 

The experimental results for the NVG transformation algorithms are shown in Fig. \ref{fig.expres_NVG_realworld} and Table \ref{table.exp_NVG_real}. In the USD\_CNY series, the proposed LOT-NVG takes $2.41E-02,\ 6.90E-02,\ 1.36E-01$ seconds to finish the transformation, which is $5.72, \ 13.85, \ 17.75$ times faster than the second-place algorithm (the DC-NVG\cite{baseline_DC}) in different tested intervals, respectively. The proposed LOT-NVG also beats all the baselines on the other two series.

The experimental results for the HVG transformation algorithms are presented in Fig. \ref{fig.expres_HVG_realworld} and Table \ref{table.exp_HVG_real}. The proposed LOT-HVG\_MSopt method achieves the best performance, with our LOT-HVG algorithm ranking second in most cases. It is worth noting that the BST-HVG method\cite{baseline_BST} is the slowest, defeated by the Basic-NVG\cite{baseline_basic} algorithm, spending more than 10 seconds to finish the transformation when the interval size is 100 in all the tested series.

\subsection{Summary}
For the NVG transformation, the LOT-NVG algorithm exhibits state-of-the-art performance in both synthetic and real-world datasets. For the HVG transformation, when the interval size exceeds 10, our LOT-HVG and LOT-HVG\_MSopt algorithms surpass the previous state-of-the-art method (the LT method\cite{baseline_MS}). Furthermore, the LT method is only applicable to the HVG transformation; in contrast, our LOT framework is suitable for both NVG and HVG transformations, demonstrating a high degree of generalization ability.

The superior performance of the proposed LOT-NVG, LOT-HVG, and LOT-HVG\_MSopt algorithms validates the effectiveness of the LOT framework.

\section{Conclusions}
\label{sec.conclu}
In this paper, we initially propose the standards of VG's online functionality by synthesizing previous studies. To accelerate the speed of VG's transformation, we introduce the LOT framework, guided by the proposed criteria, and present its implementations for the NVG(LOT-NVG) and HVG(LOT-HVG). Furthermore, we integrate the LOT-HVG with the monotonic stack concept from article\cite{baseline_MS}, resulting in the development of the LOT-HVG\_MSopt method.  The innovative aspects of the LOT framework are rooted in the specially designed adjacent data structure and the "traverse from back" strategy, which facilitate the efficient elimination and efficient update in the data stream scenario. The conducted experiments substantiate the validity and the effectiveness of the LOT framework, and its superiority compared to other baselines.

\section*{Acknowledgment}
The work is partially supported by National Natural Science Foundation of China (Grant No. TBD).

\appendix
\setcounter{table}{0}
\section{Code snippets}
\label{app.code}
\begin{lstlisting}[language=Python, firstline=1, lastline=28, label=lst.NVG, caption=Visibility checking function for NVG.]
# Python-style code snippet for demonstration.
def visibility_check_NVG(series, ticks, target_index, left_index, right_index):
    value_target = series[target_index]
    tick_target = ticks[target_index]
    nodes_visible = [] # nodes that the node with node_ID could see
    # generally, VG is in the form of undirected graph, which means that node_ID are connected to nodes_visible with undirected edges
    
    min_slope = float("inf")
    for j in range(target_index - 1, left_index - 1, -1):
        tick_j = ticks[j]
        value_j = series[j]
        slope = (value_target - value_j) * 1.0 / (tick_target - tick_j)
        if slope <= min_slope:
            if value_target >= value_j:
                nodes_visible.append(tick_j)
            min_slope = slope

    max_slope = float("-inf")
    for j in range(target_index + 1, right_index + 1):
        tick_j = ticks[j]
        value_j = series[j]
        slope = (value_j - value_target) * 1.0 / (tick_j - tick_target)
        if slope >= max_slope:
            if value_target >= value_j:
                nodes_visible.append(tick_j)
            max_slope = slope

    return (tick_target, nodes_visible)
\end{lstlisting}

\begin{lstlisting}[language=Python, firstline=1, lastline=26, label=lst.HVG, caption=Visibility checking function for HVG.]
# Python-style code snippet for demonstration.
def visibility_check_HVG(series, ticks, target_index, left_index, right_index):
    value_target = series[target_index]
    tick_target = ticks[target_index]
    nodes_visible = [] # nodes that the node of tick_target could see
    # generally, VG is in the form of undirected graph, which means that the node of tick_target are connected to nodes_visible with undirected edges

    max_height = float("-inf")
    for j in range(target_index - 1, left_index - 1, -1):
        value_j = series[j]
        tick_j = ticks[j]
        if value_j > max_height:
            if value_j <= value_target:
                nodes_visible.append(tick_j)
            max_height = value_j
    
    max_height = float("-inf")
    for j in range(target_index + 1, right_index + 1):
        value_j = series[j]
        tick_j = ticks[j]
        if value_j > max_height:
            if value_j <= value_target:
                nodes_visible.append(tick_j)
            max_height = value_j

    return (tick_target, nodes_visible)
\end{lstlisting}

\begin{lstlisting}[language=Python, firstline=1, lastline=20, label=lst.LOT-NVG, caption=Efficient update function for NVG.]
# Python-style code snippet for demonstration.
from collections import defaultdict

def efficient_update_NVG(adj_dict: defaultdict, ori_series: list, index_left: int, index_right: int) -> defaultdict:
    # Efficient update.
    value_last = ori_series[index_right]
    min_slope = float("inf")
    for node_index in range(index_right - 1, index_left - 1, -1):
        current_value = ori_series[node_index]
        slope = (value_last - current_value) * 1.0 / (index_right - node_index)
        if slope <= min_slope:
            if slope > 0:
                adj_dict[index_right].add(node_index)
            elif slope < 0:
                adj_dict[node_index].add(index_right)
            else:
                adj_dict[index_right].add(node_index)
                adj_dict[node_index].add(index_right)
            min_slope = slope
    return adj_dict
\end{lstlisting}

\begin{lstlisting}[language=Python, firstline=1, lastline=21, label=lst.LOT-HVG, caption=Efficient update function for HVG.]
# Python-style code snippet for demonstration.
from collections import defaultdict

def efficient_update_HVG(adj_dict: defaultdict, ori_series: list, index_left: int, index_right: int) -> defaultdict:
    # Efficient update.
    value_last = ori_series[index_right]
    max_value = float("-inf")
    for node_index in range(index_right - 1, index_left - 1, -1):
        current_value = ori_series[node_index]
        if max_value < value_last and current_value > max_value:
            if current_value > value_last:
                adj_dict[node_index].add(index_right)
            elif current_value < value_last:
                adj_dict[index_right].add(node_index)
            else:
                adj_dict[index_right].add(node_index)
                adj_dict[node_index].add(index_right)
            max_value = current_value
        if max_value >= value_last:
            break
    return adj_dict
\end{lstlisting}

\section{Detailed experimental data}
\label{app.data}

\begin{table}[htb]
\centering
\caption{Computational time (seconds) of different NVG transformation algorithms in synthetic dataset -- Part 1.}
\resizebox{1.05\textwidth}{11mm}{
\begin{tabular}{cccccccccc|ccccccccc}
\toprule[0.75pt]
 & \multicolumn{18}{c}{Size of interval} \\
Series & \multicolumn{9}{c}{Random series(uniform distribution)} & \multicolumn{9}{c}{Random series(normal distribution)} \\
Methods & 10 & 50 & 100 & 250 & 500 & 750 & 1000 & 1500 & 2000 & 10 & 50 & 100 & 250 & 500 & 750 & 1000 & 1500 & 2000 \\
\midrule[0.3pt]
Basic-NVG\cite{baseline_basic} & 1.19E-04 & 8.22E-04 & 3.41E-03 & 1.80E-02 & 7.14E-02 & 1.60E-01 & 2.85E-01 & 6.52E-01 & 1.15E+00 & 6.48E-05 & 8.07E-04 & 3.03E-03 & 1.74E-02 & 7.15E-02 & 1.60E-01 & 2.83E-01 & 6.40E-01 & 1.14E+00 \\
BST-NVG\cite{baseline_BST} & 1.01E-04 & 1.06E-03 & 3.02E-03 & 1.43E-02 & 5.26E-02 & 1.13E-01 & 1.95E-01 & 3.94E-01 & 6.72E-01 & 1.03E-04 & 9.45E-04 & 3.03E-03 & 1.40E-02 & 5.10E-02 & 9.50E-02 & 1.90E-01 & 4.48E-01 & 8.44E-01 \\
DC-NVG\cite{baseline_DC} & 5.27E-05 & 2.57E-04 & 5.54E-04 & 1.49E-03 & 3.43E-03 & 5.13E-03 & 6.99E-03 & 1.03E-02 & 1.40E-02 & 5.44E-05 & 2.28E-04 & 4.87E-04 & 1.43E-03 & 3.13E-03 & 4.80E-03 & 6.70E-03 & 1.04E-02 & 1.46E-02 \\
SC-NVG\cite{baseline_SC} & 5.94E-05 & 3.27E-04 & 6.51E-04 & 1.94E-03 & 3.93E-03 & 6.08E-03 & 8.17E-03 & 1.22E-02 & 1.68E-02 & 5.98E-05 & 3.11E-04 & 5.96E-04 & 1.87E-03 & 3.82E-03 & 5.88E-03 & 8.08E-03 & 1.25E-02 & 1.76E-02 \\
\textbf{LOT-NVG(Ours)} & \textbf{1.62E-05} & \textbf{3.58E-05} & \textbf{6.64E-05} & \textbf{1.42E-04} & \textbf{2.64E-04} & \textbf{4.01E-04} & \textbf{5.58E-04} & \textbf{7.79E-04} & \textbf{1.02E-03} & \textbf{1.81E-05} & \textbf{3.80E-05} & \textbf{6.60E-05} & \textbf{1.56E-04} & \textbf{2.80E-04} & \textbf{4.32E-04} & \textbf{5.15E-04} & \textbf{8.39E-04} & \textbf{1.08E-03} \\
\bottomrule[0.75pt]
\end{tabular}
\label{table.exp_NVG_syn_1}
}
\\\footnotesize{a. The bold text indicates the best performance.}
\end{table}

\begin{table}[htb]
\centering
\caption{Computational time (seconds) of different NVG transformation algorithms in synthetic dataset -- Part 2.}
\resizebox{1.05\textwidth}{11mm}{
\begin{tabular}{cccccccccc|ccccccccc}
\toprule[0.75pt]
 & \multicolumn{18}{c}{Size of interval} \\
Series & \multicolumn{9}{c}{Random series(uniform distribution)} & \multicolumn{9}{c}{Random series(normal distribution)} \\
Methods & 10 & 50 & 100 & 250 & 500 & 750 & 1000 & 1500 & 2000 & 10 & 50 & 100 & 250 & 500 & 750 & 1000 & 1500 & 2000 \\
\midrule[0.3pt]
Basic-NVG\cite{baseline_basic} & 6.26E-05 & 8.26E-04 & 3.10E-03 & 1.74E-02 & 7.00E-02 & 1.59E-01 & 2.84E-01 & 6.42E-01 & 1.14E+00 & 6.19E-05 & 8.51E-04 & 3.00E-03 & 1.74E-02 & 7.05E-02 & 1.60E-01 & 2.85E-01 & 6.44E-01 & 1.14E+00 \\
BST-NVG\cite{baseline_BST} & 9.86E-05 & 9.66E-04 & 2.96E-03 & 2.01E-02 & 5.24E-02 & 1.30E-01 & 2.86E-01 & 4.92E-01 & 8.19E-01 & 1.52E-04 & 4.63E-03 & 2.20E-02 & 1.57E-01 & 6.29E-01 & 1.07E+00 & 1.61E+00 & 4.77E+00 & 8.61E+00 \\
DC-NVG\cite{baseline_DC} & 1.46E-04 & 2.55E-04 & 5.56E-04 & 1.52E-03 & 3.20E-03 & 5.28E-03 & 7.22E-03 & 1.06E-02 & 1.45E-02 & 6.11E-05 & 5.11E-04 & 1.53E-03 & 6.74E-03 & 1.75E-02 & 2.14E-02 & 2.80E-02 & 5.75E-02 & 8.03E-02 \\
SC-NVG\cite{baseline_SC} & 5.99E-05 & 3.18E-04 & 6.70E-04 & 1.89E-03 & 3.89E-03 & 6.17E-03 & 8.74E-03 & 1.30E-02 & 1.80E-02 & 7.09E-05 & 7.22E-04 & 1.82E-03 & 6.99E-03 & 1.80E-02 & 2.29E-02 & 2.91E-02 & 6.10E-02 & 8.77E-02 \\
\textbf{LOT-NVG(Ours)} & \textbf{1.60E-05} & \textbf{4.09E-05} & \textbf{6.58E-05} & \textbf{1.46E-04} & \textbf{2.56E-04} & \textbf{3.97E-04} & \textbf{5.55E-04} & \textbf{8.02E-04} & \textbf{1.05E-03} & \textbf{2.20E-05} & \textbf{5.03E-05} & \textbf{8.38E-05} & \textbf{2.12E-04} & \textbf{3.97E-04} & \textbf{5.88E-04} & \textbf{7.45E-04} & \textbf{1.43E-03} & \textbf{1.78E-03} \\
\bottomrule[0.75pt]
\end{tabular}
\label{table.exp_NVG_syn_2}
}
\\\footnotesize{a. The bold text indicates the best performance.}
\end{table}

\begin{table}[hbt]
\centering
\caption{Computational time (seconds) of different NVG transformation algorithms in synthetic dataset -- Part 3.}
\resizebox{0.6\textwidth}{11mm}{
\begin{tabular}{cccccccccc}
\toprule[0.75pt]
 & \multicolumn{9}{c}{Size of interval} \\
Series & \multicolumn{9}{c}{Random walk series}\\
Methods & 10 & 50 & 100 & 250 & 500 & 750 & 1000 & 1500 & 2000\\
\midrule[0.3pt]
Basic-NVG\cite{baseline_basic} & 5.99E-05 & 8.23E-04 & 3.02E-03 & 1.73E-02 & 7.03E-02 & 1.60E-01 & 2.84E-01 & 6.41E-01 & 1.14E+00 \\
BST-NVG\cite{baseline_BST} & 1.08E-04 & 1.68E-03 & 5.49E-03 & 3.58E-02 & 3.28E-01 & 8.16E-01 & 1.16E+00 & 1.03E+01 & 1.48E+01 \\
DC-NVG\cite{baseline_DC} & 5.19E-05 & 4.73E-04 & 6.31E-04 & 2.37E-03 & 8.19E-03 & 1.42E-02 & 1.80E-02 & 5.57E-02 & 7.75E-02 \\
SC-NVG\cite{baseline_SC} & 6.18E-05 & 5.70E-04 & 8.73E-04 & 2.86E-03 & 9.90E-03 & 1.67E-02 & 2.17E-02 & 6.18E-02 & 8.44E-02 \\
\textbf{LOT-NVG(Ours)} & \textbf{1.40E-05} & \textbf{3.39E-05} & \textbf{7.13E-05} & \textbf{1.58E-04} & \textbf{3.59E-04} & \textbf{6.20E-04} & \textbf{6.88E-04} & \textbf{1.38E-03} & \textbf{1.76E-03}
\\
\bottomrule[0.75pt]
\end{tabular}
\label{table.exp_NVG_syn_3}
}
\\\footnotesize{a. The bold text indicates the best performance.}
\end{table}

\begin{table}[htb]
\centering
\caption{Computational time (seconds) of different HVG transformation algorithms in synthetic dataset -- Part 1.}
\resizebox{1.05\textwidth}{15mm}{
\begin{tabular}{cccccccccc|ccccccccc}
\toprule[0.75pt]
 & \multicolumn{18}{c}{Size of interval} \\
Series & \multicolumn{9}{c}{Random series(uniform distribution)} & \multicolumn{9}{c}{Random series(normal distribution)} \\
Methods & 10 & 50 & 100 & 250 & 500 & 750 & 1000 & 1500 & 2000 & 10 & 50 & 100 & 250 & 500 & 750 & 1000 & 1500 & 2000 \\
\midrule[0.3pt]
Basic-HVG\cite{baseline_basic} & 1.05E-04 & 2.50E-04 & 7.20E-04 & 3.78E-03 & 1.42E-02 & 3.19E-02 & 5.78E-02 & 1.30E-01 & 2.23E-01 & 3.98E-05 & 3.15E-04 & 7.15E-04 & 3.65E-03 & 1.39E-02 & 3.12E-02 & 5.80E-02 & 1.32E-01 & 2.33E-01 \\
BST-HVG\cite{baseline_BST} & 6.07E-05 & 3.51E-04 & 5.83E-04 & 1.68E-03 & 3.61E-03 & 5.36E-03 & 7.45E-03 & 1.18E-02 & 1.52E-02 & 5.19E-05 & 2.62E-04 & 5.65E-04 & 1.48E-03 & 3.46E-03 & 5.09E-03 & 7.75E-03 & 1.20E-02 & 1.57E-02 \\
DC-HVG\cite{baseline_DC} & 4.80E-05 & 1.81E-04 & 3.70E-04 & 1.25E-03 & 2.01E-03 & 3.07E-03 & 4.22E-03 & 6.92E-03 & 8.35E-03 & 4.19E-05 & 1.79E-04 & 3.65E-04 & 9.08E-04 & 1.90E-03 & 2.99E-03 & 4.77E-03 & 6.58E-03 & 8.45E-03 \\
SC-HVG\cite{baseline_SC} & 4.64E-05 & 2.15E-04 & 4.44E-04 & 1.14E-03 & 2.59E-03 & 3.91E-03 & 5.36E-03 & 8.01E-03 & 1.07E-02 & 4.59E-05 & 2.25E-04 & 4.46E-04 & 1.07E-03 & 2.56E-03 & 3.83E-03 & 5.18E-03 & 9.43E-03 & 1.22E-02 \\
\textbf{LOT-HVG(Ours)} & 1.89E-05 & \uline{2.52E-05} & \textbf{2.88E-05} & \uline{5.78E-05} & \uline{1.02E-04} & \uline{1.76E-04} & \uline{2.67E-04} & \uline{2.64E-04} & \uline{3.67E-04} & 1.60E-05 & \textbf{2.60E-05} & \uline{3.48E-05} & \uline{6.08E-05} & \uline{1.12E-04} & \textbf{1.39E-04} & \uline{2.00E-04} & \uline{2.85E-04} & \uline{4.51E-04} \\
LT\cite{baseline_MS} & \textbf{1.45E-05} & 3.88E-05 & 6.78E-05 & 1.56E-04 & 3.17E-04 & 4.71E-04 & 5.96E-04 & 8.94E-04 & 9.64E-04 & \textbf{1.00E-05} & 3.98E-05 & 6.64E-05 & 1.52E-04 & 2.97E-04 & 5.19E-04 & 5.51E-04 & 9.21E-04 & 9.83E-04 \\
\textbf{LOT-HVG\_MSopt(Ours)} & \uline{1.69E-05} & \textbf{2.23E-05} & \uline{3.03E-05} & \textbf{5.19E-05} & \textbf{8.18E-05} & \textbf{1.29E-04} & \textbf{1.68E-04} & \textbf{2.33E-04} & \textbf{2.88E-04} & \uline{1.59E-05} & \uline{2.79E-05} & \textbf{3.14E-05} & \textbf{5.54E-05} & \textbf{9.22E-05} & \uline{1.60E-04} & \textbf{1.61E-04} & \textbf{2.51E-04} & \textbf{2.76E-04} \\
\bottomrule[0.75pt]
\end{tabular}
\label{table.exp_HVG_syn_1}
}
\footnotesize{a. The bold text and the underlined text indicate the best \\\vspace{-3mm}and the secondary best performance respectively.}
\end{table}

\begin{table}[htb]
\centering
\caption{Computational time (seconds) of different HVG transformation algorithms in synthetic dataset -- Part 2.}
\resizebox{1.05\textwidth}{15mm}{
\begin{tabular}{cccccccccc|ccccccccc}
\toprule[0.75pt]
 & \multicolumn{18}{c}{Size of interval} \\
Series & \multicolumn{9}{c}{Random series(uniform distribution)} & \multicolumn{9}{c}{Random series(normal distribution)} \\
Methods & 10 & 50 & 100 & 250 & 500 & 750 & 1000 & 1500 & 2000 & 10 & 50 & 100 & 250 & 500 & 750 & 1000 & 1500 & 2000 \\
\midrule[0.3pt]
Basic-HVG\cite{baseline_basic} & 3.79E-05 & 2.43E-04 & 7.00E-04 & 3.74E-03 & 1.50E-02 & 3.16E-02 & 5.58E-02 & 1.29E-01 & 2.20E-01 & 4.39E-05 & 2.58E-04 & 7.74E-04 & 3.83E-03 & 1.44E-02 & 3.14E-02 & 5.58E-02 & 1.28E-01 & 2.38E-01 \\
BST-HVG\cite{baseline_BST} & 5.98E-05 & 2.76E-04 & 5.83E-04 & 2.11E-03 & 3.87E-03 & 5.31E-03 & 7.49E-03 & 1.10E-02 & 1.49E-02 & 5.78E-05 & 4.58E-04 & 1.35E-03 & 6.09E-03 & 1.56E-02 & 1.92E-02 & 2.51E-02 & 5.00E-02 & 7.07E-02 \\
DC-HVG\cite{baseline_DC} & 4.20E-05 & 1.92E-04 & 3.58E-04 & 1.13E-03 & 2.30E-03 & 3.06E-03 & 4.27E-03 & 6.34E-03 & 8.26E-03 & 4.49E-05 & 2.30E-04 & 5.68E-04 & 2.04E-03 & 5.47E-03 & 7.08E-03 & 9.33E-03 & 1.77E-02 & 2.53E-02 \\
SC-HVG\cite{baseline_SC} & 4.79E-05 & 2.13E-04 & 4.32E-04 & 1.28E-03 & 2.55E-03 & 3.87E-03 & 5.41E-03 & 8.45E-03 & 1.04E-02 & 4.73E-05 & 2.65E-04 & 6.00E-04 & 2.01E-03 & 6.15E-03 & 1.20E-02 & 1.97E-02 & 5.61E-02 & 9.45E-02 \\
\textbf{LOT-HVG(Ours)} & 1.60E-05 & \textbf{2.00E-05} & \uline{3.26E-05} & 6.14E-05 & \uline{1.04E-04} & \uline{1.34E-04} & \uline{1.92E-04} & \uline{3.20E-04} & \uline{3.42E-04} & \textbf{1.20E-05} & \uline{1.30E-04} & \uline{3.56E-05} & \uline{5.78E-05} & \uline{1.06E-04} & \uline{1.45E-04} & \uline{1.85E-04} & \uline{2.91E-04} & \uline{3.33E-04} \\
LT\cite{baseline_MS} & \uline{1.40E-05} & 3.60E-05 & 6.59E-05 & \uline{1.55E-04} & 2.87E-04 & 4.87E-04 & 5.50E-04 & 9.02E-04 & 9.31E-04 & \textbf{1.20E-05} & 5.59E-05 & 5.55E-05 & 1.17E-04 & 2.47E-04 & 3.63E-04 & 3.96E-04 & 7.13E-04 & 8.30E-04 \\
\textbf{LOT-HVG\_MSopt(Ours)} & \textbf{1.20E-05} & \uline{2.39E-05} & \textbf{3.02E-05} & \textbf{5.17E-05} & \textbf{9.12E-05} & \textbf{1.21E-04} & \textbf{1.51E-04} & \textbf{2.32E-04} & \textbf{2.76E-04} & 1.39E-05 & \textbf{2.19E-05} & \textbf{3.24E-05} & \textbf{5.38E-05} & \textbf{9.52E-05} & \textbf{1.28E-04} & \textbf{1.51E-04} & \textbf{2.51E-04} & \textbf{2.83E-04} \\
\bottomrule[0.75pt]
\end{tabular}
\label{table.exp_HVG_syn_2}
}
\footnotesize{a. The bold text and the underlined text indicate the best \\\vspace{-3mm}and the secondary best performance respectively.}
\end{table}

\begin{table}[hbt]
\centering
\caption{Computational time (seconds) of different HVG transformation algorithms in synthetic dataset -- Part 3.}
\resizebox{0.7\textwidth}{15mm}{
\begin{tabular}{cccccccccc}
\toprule[0.75pt]
 & \multicolumn{9}{c}{Size of interval} \\
Series & \multicolumn{9}{c}{Random walk series}\\
Methods & 10 & 50 & 100 & 250 & 500 & 750 & 1000 & 1500 & 2000\\
\midrule[0.3pt]
Basic-HVG\cite{baseline_basic} & 3.40E-05 & 2.14E-04 & 6.90E-04 & 3.45E-03 & 1.39E-02 & 3.09E-02 & 5.50E-02 & 1.25E-01 & 2.20E-01 \\
BST-HVG\cite{baseline_BST} & 5.00E-05 & 2.67E-04 & 6.54E-04 & 2.14E-03 & 7.43E-03 & 1.27E-02 & 1.60E-02 & 4.88E-02 & 6.83E-02 \\
DC-HVG\cite{baseline_DC} & 3.95E-05 & 1.74E-04 & 3.81E-04 & 1.05E-03 & 2.99E-03 & 5.12E-03 & 6.64E-03 & 1.67E-02 & 2.34E-02 \\
SC-HVG\cite{baseline_SC} & 4.46E-05 & 2.36E-04 & 5.82E-04 & 2.16E-03 & 7.98E-03 & 1.46E-02 & 2.35E-02 & 5.01E-02 & 7.54E-02 \\
\textbf{LOT-HVG(Ours)} & \uline{1.53E-05} & \uline{2.61E-05} & \uline{3.19E-05} & \uline{6.12E-05} & \uline{1.07E-04} & \uline{1.40E-04} & \uline{1.84E-04} & \uline{3.05E-04} & \uline{3.71E-04} \\
LT\cite{baseline_MS} & 1.54E-05 & 3.67E-05 & 5.78E-05 & 1.23E-04 & 2.67E-04 & 3.82E-04 & 4.80E-04 & 9.65E-04 & 9.42E-04 \\
\textbf{LOT-HVG\_MSopt(Ours)} & \textbf{1.41E-05} & \textbf{2.41E-05} & \textbf{2.79E-05} & \textbf{5.33E-05} & \textbf{9.53E-05} & \textbf{1.31E-04} & \textbf{1.67E-04} & \textbf{2.89E-04} & \textbf{3.09E-04}
\\
\bottomrule[0.75pt]
\end{tabular}
\label{table.exp_HVG_syn_3}
}
\\\footnotesize{a. The bold text and the underlined text indicate the best \\\vspace{-3mm}and the secondary best performance respectively.}
\end{table}

\begin{table}[hbt]
\centering
\caption{Computational time (seconds) of different NVG transformation algorithms in real-world dataset.}
\resizebox{0.7\textwidth}{13mm}{
\begin{tabular}{cccccccccc}
\toprule[0.75pt]
 & \multicolumn{9}{c}{Size of interval} \\
Series & \multicolumn{3}{c}{USD\_CNY} & \multicolumn{3}{c}{VIX} & \multicolumn{3}{c}{WTI}\\
Methods & 10 & 50 & 100 & 10 & 50 & 100 & 10 & 50 & 100\\
\midrule[0.3pt]
Basic-NVG\cite{baseline_basic} & 1.26E-01 & 2.01E+00 & 7.75E+00 & 1.12E-01 & 1.90E+00 & 7.18E+00 & 1.13E-01 & 1.97E+00 & 7.26E+00 \\
BST-NVG\cite{baseline_BST} & 2.99E-01 & 5.77E+00 & 2.55E+01 & 2.80E-01 & 4.74E+00 & 1.82E+01 & 2.97E-01 & 5.28E+00 & 2.18E+01 \\
DC-NVG\cite{baseline_DC} & 9.70E-02 & 7.65E-01 & 1.92E+00 & 9.15E-02 & 6.42E-01 & 1.60E+00 & 9.36E-02 & 7.11E-01 & 1.84E+00 \\
SC-NVG\cite{baseline_SC} & 1.38E-01 & 9.56E-01 & 2.41E+00 & 1.15E-01 & 9.08E-01 & 2.08E+00 & 1.15E-01 & 8.63E-01 & 2.28E+00 \\
\textbf{LOT-NVG(Ours)} & \textbf{2.41E-02} & \textbf{6.90E-02} & \textbf{1.36E-01} & \textbf{1.89E-02} & \textbf{6.80E-02} & \textbf{1.29E-01} & \textbf{1.89E-02} & \textbf{6.67E-02} & \textbf{1.33E-01}
\\
\bottomrule[0.75pt]
\end{tabular}
\label{table.exp_NVG_real}
}
\\\footnotesize{a. The bold text indicates the best performance.}
\end{table}

\begin{table}[hbt]
\centering
\caption{Computational time (seconds) of different HVG transformation algorithms in real-world dataset.}
\resizebox{0.7\textwidth}{15mm}{
\begin{tabular}{cccccccccc}
\toprule[0.75pt]
 & \multicolumn{9}{c}{Size of interval} \\
Series & \multicolumn{3}{c}{USD\_CNY} & \multicolumn{3}{c}{VIX} & \multicolumn{3}{c}{WTI}\\
Methods & 10 & 50 & 100 & 10 & 50 & 100 & 10 & 50 & 100\\
\midrule[0.3pt]
Basic-HVG\cite{baseline_basic} & 1.63E-01 & 1.99E+00 & 7.21E+00 & 1.44E-01 & 1.87E+00 & 6.76E+00 & 1.48E-01 & 2.01E+00 & 7.19E+00 \\
BST-HVG\cite{baseline_BST} & 5.16E-01 & 3.96E+00 & 1.07E+01 & 5.08E-01 & 3.57E+00 & 9.18E+00 & 5.22E-01 & 4.06E+00 & 1.02E+01 \\
DC-HVG\cite{baseline_DC} & 2.13E-01 & 1.23E+00 & 3.00E+00 & 2.09E-01 & 1.15E+00 & 2.72E+00 & 2.17E-01 & 1.24E+00 & 2.90E+00 \\
SC-HVG\cite{baseline_SC} & 2.10E-01 & 1.33E+00 & 3.17E+00 & 2.06E-01 & 1.22E+00 & 2.73E+00 & 2.11E-01 & 1.27E+00 & 3.04E+00 \\
\textbf{LOT-HVG(Ours)} & \uline{3.81E-02} & \uline{8.09E-02} & \uline{1.34E-01} & \uline{3.60E-02} & \uline{7.54E-02} & \uline{1.25E-01} & 4.36E-02 & \uline{8.90E-02} & \uline{1.40E-01} \\
LT\cite{baseline_MS} & 3.87E-02 & 1.40E-01 & 2.71E-01 & 3.85E-02 & 1.43E-01 & 2.78E-01 & \uline{3.99E-02} & 1.45E-01 & 2.86E-01 \\
\textbf{LOT-HVG\_MSopt(Ours)} & \textbf{3.39E-02} & \textbf{6.98E-02} & \textbf{1.13E-01} & \textbf{3.26E-02} & \textbf{6.50E-02} & \textbf{1.08E-01} & \textbf{3.74E-02} & \textbf{6.86E-02} & \textbf{1.12E-01}
\\
\bottomrule[0.75pt]
\end{tabular}
\label{table.exp_HVG_real}
}
\\\footnotesize{a. The bold text and the underlined text indicate the best \\\vspace{-3mm}and the secondary best performance respectively.}
\end{table}




\bibliographystyle{elsarticle-num-names} 
\bibliography{LOT_algo.bib}

\begin{thebibliography}{21}
\expandafter\ifx\csname natexlab\endcsname\relax\def\natexlab#1{#1}\fi
\providecommand{\url}[1]{\texttt{#1}}
\providecommand{\href}[2]{#2}
\providecommand{\path}[1]{#1}
\providecommand{\DOIprefix}{doi:}
\providecommand{\ArXivprefix}{arXiv:}
\providecommand{\URLprefix}{URL: }
\providecommand{\Pubmedprefix}{pmid:}
\providecommand{\doi}[1]{\href{http://dx.doi.org/#1}{\path{#1}}}
\providecommand{\Pubmed}[1]{\href{pmid:#1}{\path{#1}}}
\providecommand{\bibinfo}[2]{#2}
\ifx\xfnm\relax \def\xfnm[#1]{\unskip,\space#1}\fi
\bibitem[{Lacasa et~al.(2008)Lacasa, Luque, Ballesteros, Luque, and Nuno}]{GG_NVG}
\bibinfo{author}{L.~Lacasa}, \bibinfo{author}{B.~Luque}, \bibinfo{author}{F.~Ballesteros}, \bibinfo{author}{J.~Luque}, \bibinfo{author}{J.~C. Nuno},
\newblock \bibinfo{title}{From time series to complex networks: The visibility graph},
\newblock \bibinfo{journal}{Proceedings of the National Academy of Sciences} \bibinfo{volume}{105} (\bibinfo{year}{2008}) \bibinfo{pages}{4972--4975}.
\bibitem[{Luque et~al.(2009)Luque, Lacasa, Ballesteros, and Luque}]{GG_HVG}
\bibinfo{author}{B.~Luque}, \bibinfo{author}{L.~Lacasa}, \bibinfo{author}{F.~Ballesteros}, \bibinfo{author}{J.~Luque},
\newblock \bibinfo{title}{Horizontal visibility graphs: Exact results for random time series},
\newblock \bibinfo{journal}{Physical Review E} \bibinfo{volume}{80} (\bibinfo{year}{2009}) \bibinfo{pages}{046103}.
\bibitem[{Bezsudnov and Snarskii(2014)}]{GG_PNVG}
\bibinfo{author}{I.~Bezsudnov}, \bibinfo{author}{A.~Snarskii},
\newblock \bibinfo{title}{From the time series to the complex networks: The parametric natural visibility graph},
\newblock \bibinfo{journal}{Physica A: Statistical Mechanics and its Applications} \bibinfo{volume}{414} (\bibinfo{year}{2014}) \bibinfo{pages}{53--60}.
\bibitem[{Gao et~al.(2016)Gao, Cai, Yang, Dang, and Zhang}]{GG_LPHVG}
\bibinfo{author}{Z.-K. Gao}, \bibinfo{author}{Q.~Cai}, \bibinfo{author}{Y.-X. Yang}, \bibinfo{author}{W.-D. Dang}, \bibinfo{author}{S.-S. Zhang},
\newblock \bibinfo{title}{Multiscale limited penetrable horizontal visibility graph for analyzing nonlinear time series},
\newblock \bibinfo{journal}{Scientific reports} \bibinfo{volume}{6} (\bibinfo{year}{2016}) \bibinfo{pages}{1--7}.
\bibitem[{Wang et~al.(2018)Wang, Vilela, Du, Zhao, Dong, Tian, and Stanley}]{GG_LPHVGF}
\bibinfo{author}{M.~Wang}, \bibinfo{author}{A.~L. Vilela}, \bibinfo{author}{R.~Du}, \bibinfo{author}{L.~Zhao}, \bibinfo{author}{G.~Dong}, \bibinfo{author}{L.~Tian}, \bibinfo{author}{H.~E. Stanley},
\newblock \bibinfo{title}{Topological properties of the limited penetrable horizontal visibility graph family},
\newblock \bibinfo{journal}{Physical Review E} \bibinfo{volume}{97} (\bibinfo{year}{2018}) \bibinfo{pages}{052117}.
\bibitem[{Wen et~al.(2022)Wen, Chen, and Cheong}]{survey_VG}
\bibinfo{author}{T.~Wen}, \bibinfo{author}{H.~Chen}, \bibinfo{author}{K.~H. Cheong},
\newblock \bibinfo{title}{Visibility graph for time series prediction and image classification: a review},
\newblock \bibinfo{journal}{Nonlinear Dynamics} \bibinfo{volume}{110} (\bibinfo{year}{2022}) \bibinfo{pages}{2979--2999}.
\bibitem[{Hu et~al.(2022)Hu, Zhang, Wu, and Li}]{exa_analysis1}
\bibinfo{author}{J.~Hu}, \bibinfo{author}{Y.~Zhang}, \bibinfo{author}{P.~Wu}, \bibinfo{author}{H.~Li},
\newblock \bibinfo{title}{An analysis of the global fuel-trading market based on the visibility graph approach},
\newblock \bibinfo{journal}{Chaos, Solitons \& Fractals} \bibinfo{volume}{154} (\bibinfo{year}{2022}) \bibinfo{pages}{111613}.
\bibitem[{Sulaimany and Mafakheri(2023)}]{exa_analysis2}
\bibinfo{author}{S.~Sulaimany}, \bibinfo{author}{A.~Mafakheri},
\newblock \bibinfo{title}{Visibility graph analysis of web server log files},
\newblock \bibinfo{journal}{Physica A: Statistical Mechanics and its Applications} \bibinfo{volume}{611} (\bibinfo{year}{2023}) \bibinfo{pages}{128448}.
\bibitem[{Wang et~al.(2022)Wang, Han, Chen, Bi, Guan, and Zou}]{exa_analysis3}
\bibinfo{author}{X.~Wang}, \bibinfo{author}{X.~Han}, \bibinfo{author}{Z.~Chen}, \bibinfo{author}{Q.~Bi}, \bibinfo{author}{S.~Guan}, \bibinfo{author}{Y.~Zou},
\newblock \bibinfo{title}{Multi-scale transition network approaches for nonlinear time series analysis},
\newblock \bibinfo{journal}{Chaos, Solitons \& Fractals} \bibinfo{volume}{159} (\bibinfo{year}{2022}) \bibinfo{pages}{112026}.
\bibitem[{Hu and Xiao(2022{\natexlab{a}})}]{exa_forecast1}
\bibinfo{author}{Y.~Hu}, \bibinfo{author}{F.~Xiao},
\newblock \bibinfo{title}{An efficient forecasting method for time series based on visibility graph and multi-subgraph similarity},
\newblock \bibinfo{journal}{Chaos, Solitons \& Fractals} \bibinfo{volume}{160} (\bibinfo{year}{2022}{\natexlab{a}}) \bibinfo{pages}{112243}.
\bibitem[{Hu and Xiao(2022{\natexlab{b}})}]{exa_forecast2}
\bibinfo{author}{Y.~Hu}, \bibinfo{author}{F.~Xiao},
\newblock \bibinfo{title}{A novel method for forecasting time series based on directed visibility graph and improved random walk},
\newblock \bibinfo{journal}{Physica A: Statistical Mechanics and its Applications} \bibinfo{volume}{594} (\bibinfo{year}{2022}{\natexlab{b}}) \bibinfo{pages}{127029}.
\bibitem[{Xuan et~al.(2022)Xuan, Zhou, Qiu, Chen, Xu, Zheng, and Yang}]{exa_forecast3}
\bibinfo{author}{Q.~Xuan}, \bibinfo{author}{J.~Zhou}, \bibinfo{author}{K.~Qiu}, \bibinfo{author}{Z.~Chen}, \bibinfo{author}{D.~Xu}, \bibinfo{author}{S.~Zheng}, \bibinfo{author}{X.~Yang},
\newblock \bibinfo{title}{Avgnet: Adaptive visibility graph neural network and its application in modulation classification},
\newblock \bibinfo{journal}{IEEE Transactions on Network Science and Engineering} \bibinfo{volume}{9} (\bibinfo{year}{2022}) \bibinfo{pages}{1516--1526}.
\bibitem[{Schmidt and K{\"o}hne(2023)}]{baseline_MS}
\bibinfo{author}{J.~Schmidt}, \bibinfo{author}{D.~K{\"o}hne},
\newblock \bibinfo{title}{A simple scalable linear time algorithm for horizontal visibility graphs},
\newblock \bibinfo{journal}{Physica A: Statistical Mechanics and its Applications} \bibinfo{volume}{616} (\bibinfo{year}{2023}) \bibinfo{pages}{128601}.
\bibitem[{Yela et~al.(2020)Yela, Thalmann, Nicosia, Stowell, and Sandler}]{baseline_BST}
\bibinfo{author}{D.~F. Yela}, \bibinfo{author}{F.~Thalmann}, \bibinfo{author}{V.~Nicosia}, \bibinfo{author}{D.~Stowell}, \bibinfo{author}{M.~Sandler},
\newblock \bibinfo{title}{Online visibility graphs: Encoding visibility in a binary search tree},
\newblock \bibinfo{journal}{Physical Review Research} \bibinfo{volume}{2} (\bibinfo{year}{2020}) \bibinfo{pages}{023069}.
\bibitem[{Li and Karahanna(2015)}]{exa_recommendsys}
\bibinfo{author}{S.~S. Li}, \bibinfo{author}{E.~Karahanna},
\newblock \bibinfo{title}{Online recommendation systems in a b2c e-commerce context: a review and future directions},
\newblock \bibinfo{journal}{Journal of the association for information systems} \bibinfo{volume}{16} (\bibinfo{year}{2015}) \bibinfo{pages}{2}.
\bibitem[{Bl{\'a}zquez-Garc{\'\i}a et~al.(2021)Bl{\'a}zquez-Garc{\'\i}a, Conde, Mori, and Lozano}]{exa_anolydetect}
\bibinfo{author}{A.~Bl{\'a}zquez-Garc{\'\i}a}, \bibinfo{author}{A.~Conde}, \bibinfo{author}{U.~Mori}, \bibinfo{author}{J.~A. Lozano},
\newblock \bibinfo{title}{A review on outlier/anomaly detection in time series data},
\newblock \bibinfo{journal}{ACM Computing Surveys (CSUR)} \bibinfo{volume}{54} (\bibinfo{year}{2021}) \bibinfo{pages}{1--33}.
\bibitem[{Lacasa et~al.(2009)Lacasa, Luque, Luque, and Nuno}]{baseline_basic}
\bibinfo{author}{L.~Lacasa}, \bibinfo{author}{B.~Luque}, \bibinfo{author}{J.~Luque}, \bibinfo{author}{J.~C. Nuno},
\newblock \bibinfo{title}{The visibility graph: A new method for estimating the hurst exponent of fractional brownian motion},
\newblock \bibinfo{journal}{Europhysics Letters} \bibinfo{volume}{86} (\bibinfo{year}{2009}) \bibinfo{pages}{30001}.
\bibitem[{Ghosh and Dutta(2019)}]{baseline_SC}
\bibinfo{author}{S.~Ghosh}, \bibinfo{author}{A.~Dutta},
\newblock \bibinfo{title}{An efficient non-recursive algorithm for transforming time series to visibility graph},
\newblock \bibinfo{journal}{Physica A: Statistical Mechanics and its Applications} \bibinfo{volume}{514} (\bibinfo{year}{2019}) \bibinfo{pages}{189--202}.
\bibitem[{Lan et~al.(2015)Lan, Mo, Chen, Liu, and Deng}]{baseline_DC}
\bibinfo{author}{X.~Lan}, \bibinfo{author}{H.~Mo}, \bibinfo{author}{S.~Chen}, \bibinfo{author}{Q.~Liu}, \bibinfo{author}{Y.~Deng},
\newblock \bibinfo{title}{Fast transformation from time series to visibility graphs},
\newblock \bibinfo{journal}{Chaos: An Interdisciplinary Journal of Nonlinear Science} \bibinfo{volume}{25} (\bibinfo{year}{2015}) \bibinfo{pages}{083105}.
\bibitem[{Stephen(2021)}]{baseline_DTHVG}
\bibinfo{author}{C.~Stephen},
\newblock \bibinfo{title}{A scalable linear-time algorithm for horizontal visibility graph construction over long sequences},
\newblock in: \bibinfo{booktitle}{2021 IEEE International Conference on Big Data (Big Data)}, \bibinfo{organization}{IEEE}, \bibinfo{year}{2021}, pp. \bibinfo{pages}{40--50}.
\bibitem[{Carmona-Cabezas et~al.(2019)Carmona-Cabezas, G{\'o}mez-G{\'o}mez, Guti{\'e}rrez~de Rav{\'e}, and Jim{\'e}nez-Hornero}]{baseline_others_sliding}
\bibinfo{author}{R.~Carmona-Cabezas}, \bibinfo{author}{J.~G{\'o}mez-G{\'o}mez}, \bibinfo{author}{E.~Guti{\'e}rrez~de Rav{\'e}}, \bibinfo{author}{F.~J. Jim{\'e}nez-Hornero},
\newblock \bibinfo{title}{A sliding window-based algorithm for faster transformation of time series into complex networks},
\newblock \bibinfo{journal}{Chaos: An Interdisciplinary Journal of Nonlinear Science} \bibinfo{volume}{29} (\bibinfo{year}{2019}).

\end{thebibliography}
\end{document}